\newcommand{\cut}[1]{}
\newcommand{\KEEPNOTE}[1]{}

\newcommand{\<}{\left\langle}
\renewcommand{\>}{\right\rangle}

\def\factorial{!}
\def\rmd{{d}}
\def\rmD{{\mathbf D}}
\def\rmi{{i}}

\def\Ham{\mathcal{H}}

\def\quenched{{\cal Q}}
\def\ensemble{{\cal E}}

\def\partitionfunction{Z}

\def\localreplicaprobability{{\cal P}}

\def\safed{f_\ensemble}

\def\repZ{{\langle \partitionfunction^n\rangle_{\quenched}}}
\def\Gone{{\cal G}_1}
\def\Gtwo{{\cal G}_2}
\def\Gthree{{\cal G}_3}

\def\mB{\mathbb{B}}
\def\SpinGlassSusceptibility{{\chi_{SG}}}
\def\Susceptibility{{\chi_{Lin}}}

\def\Extr{{\mathrm{Extr}}}

\def\atanh{\mathrm{atanh}}
\def\arbitraryfunction{G}

\def\modulationsymbol{V}
\def\sparsematrixsymbol{A}
\def\couplingsymbol{J}

\def\randomfield{z}

\newcommand{\realvectornotation}[1]{\vec{#1}}

\def\vC{{\realvectornotation{C}}}

\def\vZ{{\realvectornotation{Z}}}

\def\vtau{{\realvectornotation{\tau}}}
\def\vsigma{{\realvectornotation{\sigma}}}

\def\vones{\realvectornotation{1}}
\def\vzeros{\realvectornotation{0}}
\def\vx{\realvectornotation{x}}
\def\vX{\realvectornotation{X}}

\def\vzeta{\realvectornotation{\zeta}}
\def\veta{\realvectornotation{\eta}}
\def\vsigma{\realvectornotation{\sigma}}
\def\vtau{\realvectornotation{\tau}}
\def\vS{\realvectornotation{S}}
\def\vrandomfield{\realvectornotation{z}}

\newcommand{\replicavectornotation}[1]{{\hbox{\boldmath{$#1$}}}}
\def\rvlambda{{\replicavectornotation{\lambda}}}
\def\rvsigma{{\replicavectornotation{\sigma}}}
\def\rvtau{{\replicavectornotation{\tau}}}
\def\rvS{{{\replicavectornotation{S}}}}

\def\mJ{{{\mathbb \couplingsymbol}}}
\def\mA{{\mathbb \sparsematrixsymbol}}
\def\mxi{{\mathbb \modulationsymbol}}

\newcommand{\intZ}[1]{\oint {\rmD_{#1} Z_k}}

\newcommand{\intY}[1]{\oint {\rmD_{#1} Y_\mu}}

\def\GENOP{{\Phi}}
\def\GENOPconj{{{\hat \GENOP}}}

\def\RSOP{{\pi}}

\def\Histogram{W}

\def\ij{{\langle i j \rangle}}

\newcommand{\orderedtwo}[1]{\<#1_1,#1_2\>}
\newcommand{\orderedthree}[1]{\<#1_1,#1_2,#1_3\>}
\newcommand{\orderedfour}[1]{\<#1_1,#1_2,#1_3,#1_4\>}
\newcommand{\orderedL}[1]{{\langle {#1}_1, \ldots, {#1}_{L} \rangle}}

\def\alal{{\orderedtwo{\alpha}}}
\def\alalal{{\orderedthree{\alpha}}}
\def\alalalal{{\orderedfour{\alpha}}}
\def\qal{{q_\alpha}}
\def\qalal{q_\alal}

\def\hq{{\hat q}}
\def\bq{{\bar q}}
\def\qhal{{\hq_\alpha}}
\def\qbal{{{\bq}_\alpha}}

\def\qhalal{{\hq_\alal}}
\def\qbalal{{\bq_\alal}}

\def\qbalalal{{\bq_\alalal}}
\def\qbalalalal{{\bq_\alalalal}}

\newcommand\UPDATEBIBYES[1]{}
\newcommand\UPDATEBIBNO[1]{#1}
\documentclass[aps,pre,preprint,groupedaddress]{revtex4}

\usepackage{amssymb}
\usepackage{amsmath}
\usepackage{graphicx}

\begin{document}



\title{Equilibrium properties of disordered spin models with two scale interactions}

\author{Jack Raymond}
\email[]{jack.raymond@physics.org}
\homepage[]{http://ihome.ust.hk/~jraymond}
\altaffiliation{Department of Physics,
The Hong Kong University of Science and Technology,
Clear Water Bay, 
Hong Kong}
\author{David Saad}
\email[]{saadd@aston.ac.uk}
\affiliation{Neural Computing Research Group,\\
Aston University,\\
Birmingham,\\
UK}


\date{\today}

\begin{abstract}
Methods for understanding classical disordered spin systems with interactions conforming to some idealized graphical structure are well developed. The equilibrium properties of the Sherrington-Kirkpatrick model, which has a densely connected structure, have become well understood. Many features generalize to  sparse Erdos-Renyi graph structures above the percolation threshold, and to Bethe lattices when appropriate boundary conditions apply. In this paper we consider spin states subject to a combination of sparse strong
interactions with weak dense interactions, which we term a composite model. The equilibrium properties are examined through the replica method, with exact analysis of the high temperature paramagnetic, spin glass and ferromagnetic phases by perturbative schemes. We present results of a replica symmetric variational approximations where perturbative approaches fail at lower temperature. Results demonstrate novel reentrant behaviors from spin glass to ferromagnetic phases as temperature is lowered, including transitions from replica symmetry broken to replica symmetric phases. The nature of high temperature transitions is found to be sensitive to the connectivity profile in the sparse sub-graph, with regular connectivity a discontinuous transition from the paramagnetic to ferromagnetic phases is apparent.

\end{abstract}


\maketitle

\section{\label{composite.Introduction} Introduction}

Statistical physics methods for studying disordered spin systems have become well developed. Much of the development can be traced back to early work on mean-field models for disordered magnetic systems and the theory was strongly developed in spin-glass models~\cite{Fischer:SG,Mezard:SGT}. One problem in studying spin glasses and disordered media has been in appropriately modeling the inhomogeneity within tractable frameworks. Statistical descriptions of inhomogeneity are often realized by random coupling ensembles. Small systems described in this way may have strongly varying properties, but the ensemble may be chosen so that the macroscopic description is asymptotically well defined.

\begin{figure}[htb]
\centering{
\includegraphics[width=1.0\linewidth]{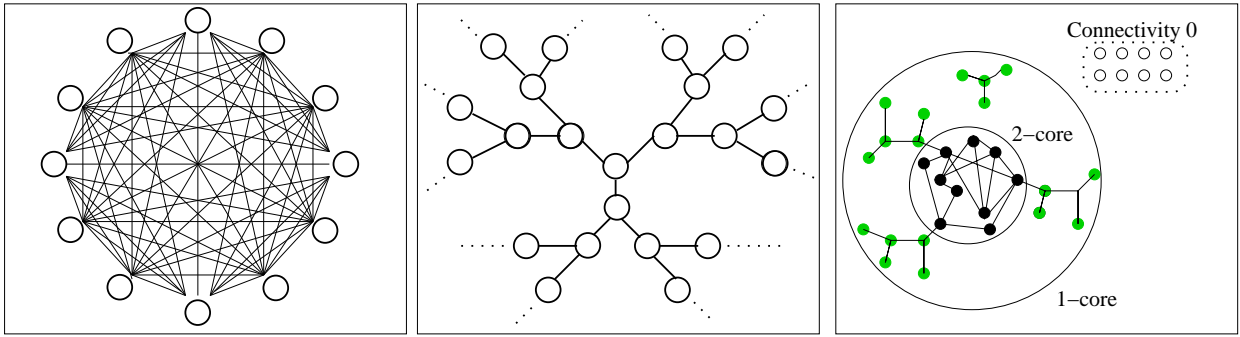}
\caption[Sparse and dense spin models.] {(color online) \label{fig:composite.2cores} Shown are the couplings (links) amongst a set of spin variables (circles), which describes graphically a particular quadratic Hamiltonian. Left figure: The fully connected graph is a special
case of a dense graph describing the SK model, with $O(N)$ non-zero couplings per variable in the large system limit. Centre/Right figure: The VB model is defined with $O(1)$ non-zero couplings per variable in the large system limit. Center figure: In the
DS - removed ','
case of a regular connectivity random graph above the percolation threshold there is an inhomogeneous structure on a global scale, but locally the structure is a Bethe lattice (regular tree). Right figure: In the case of a random graph with Poissonian connectivity the local structure is again tree like. Above the percolation threshold many trees of finite size, and unconnected variables exist, as well as a giant component containing $O(N)$ variables, and many loops~\cite{Bollobas:RG}. The 1-core contains all variables with at least one link, including the giant component above the percolation threshold.
Additional structures within the giant component may be identified, including a 2-core, obtained by recursively removing leaves (singly connected variables) from the giant component.}
}
\end{figure}

Both dense and sparse graphical models are useful in understanding a range of phenomena, such as neural networks~\cite{Hertz:ITNC}, information theory~\cite{Richardson:MCT} and other information processing~\cite{Nishimori:SP}, where spatial and dimensional constraints are often less rigid. Many complex systems have an inhomogeneous interaction structure that can be approached, if not exactly represented, by consideration of simple
random graph ensembles. In this paper, spin glass models with couplings conforming to infinite dimensional Erd\"{o}s-R\'{e}nyi random graphs are considered~\cite{Bollobas:RG}. In the large system limit many equilibrium properties depend on the connectivity distribution, and how the number of couplings per variable scales with $N$, the system size.
Dense graphs have a number of links per variable that is $O(N)$ in the large system limit, whereas sparse ensembles have finite mean connectivity in this limit. Many topological features become well defined in these limits. Two standard sparse coupling distributions are considered, a description with regular user connectivity, and one with Poissonian user connectivity. The distinctions between these two sparse models and the limiting case of full connectivity are illustrated in figure~\ref{fig:composite.2cores}.

Some densely connected models may be analyzed exactly for ensembles of uniform binary interactions, and certain random coupling models, most famously the Sherrington-Kirkpatrick (SK) model of spin glasses~\cite{Sherrington:SMSG}. Simplification of the analysis in the disordered case is often possible through noting the ability to describe large sets of interactions by central limit theorems~\cite{Ellis:ELD}. For sparse connectivity models, such as the Viana-Bray (VB) model~\cite{Viana:PD,Mottishaw:RSB}, a locally tree like approximation (Bethe
approximation) is often essential in simplifying analysis; central limit theorems again apply to certain objects, but not directly to the set of local interactions for any variable. Models which do not allow use of central limit theorems or locally tree-like approximations are normally significantly more difficult to analyze.

\cut{
Some densely connected models may be analyzed exactly for ensembles of uniform binary interactions, and certain random coupling models, most famously the Sherrington-Kirkpatrick (SK) model of spin glasses~\cite{Sherrington:SMSG}. Simplification of the analysis in the disordered case is often possible through noting the ability to describe large sets of interactions by central limit theorems~\cite{Ellis:ELD}. For sparse graphical models a locally tree like approximation (Bethe approximation) is often essential in simplifying analysis, central limit theorems again apply to certain objects, but not directly to the set of local interactions for any variable. Models which do not allow use of central limit theorems or locally tree-like approximations are normally significantly more difficult to analyze.}

Frameworks in which an interplay between strong sparse and weak pervasive couplings might be proposed in a variety of areas. In nanotechnology for example, miniaturization of classical components will preserve engineered short range interactions, but other accidental correlations may emerge not limited by the designed connectivity structure, and these may well be modeled by a mean-field (infinite connectivity) like interaction. A mixed connectivity may also be a designed feature. Neuronal activity is known to involve a combination short and long range information processing structures, this motivated a $1+\infty$ dimensional model of neuronal activity~\cite{Skantzos:1ID} discovering many novel properties. Another example of such an engineering application is CDMA, where improvement over standard methods is possible~\cite{Raymond:CC}.

To motivate a closely related study, Hase and Mendes noted a possible application for theories of these structures~\cite{Hase:DA}: Consider the model with sparse anti-ferromagnetic couplings on a structure otherwise fully connected through ferromagnetic couplings. This composite model can be considered as one in which a ferromagnetic phase is maintained by a densely connected network, but with a small proportion of links attacked. Often only a small portion of a link structure is accessible to an attacker, so it is interesting to consider how the system response differs from weak attacks on all (or most) links.

The effect of an attack on a sparse subset may cause a transition away from the ordered phase, when sufficiently strong. It is possible that the nature of transitions away from the ordered state may differ from those with only a single interaction scale. The effect of disruption of networks by random attack, or frustrating interactions, is of importance in many practical network models~\cite{Albert:EA,Hase:DA}, the restriction to random topologies allows a focus on generic properties, in this case restricted to the issue of sparse and dense induced effects.

\begin{figure}[htb]
\centering{
\includegraphics[width=0.8\linewidth]{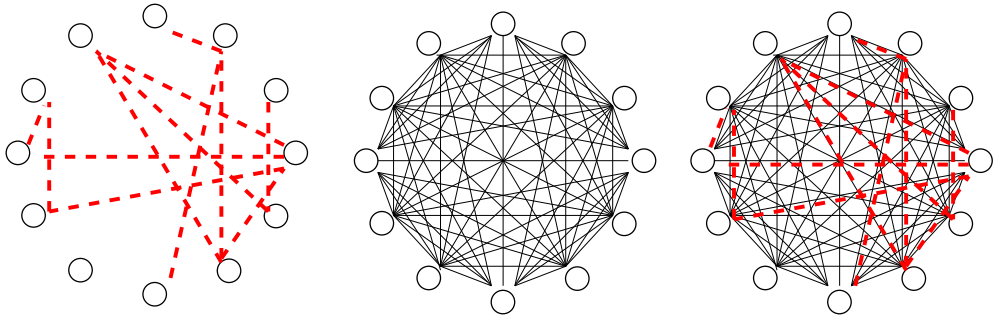}
\caption[Composite models.]{(color online) \label{fig:composite.COMPSYS} Left figure: A sparse model defined by some mean connectivity, describes couplings in the sparse model. Center figure: A fully connected model, describes couplings in the dense model. Right figure: A fully connected graph with a subset of strong sparse links, this is the composite model. The sparse subset of couplings are an order of magnitude stronger than the couplings on the other edges.}
}
\end{figure}

More generally, a range of mean field behavior, including spin glass like, may be supported by the dense sub-structure, combined with an arbitrary set supported by the
sparse sub-structure, as shown in figure~\ref{fig:composite.COMPSYS}. In so doing, a wider variety of competitive phase behavior are explored.

It may be expected that many of the results for composite systems will be similar to those for the limiting sparse and dense models. Four thermodynamic phases describe equilibrium properties of spin models with independent and identically distributed (i.i.d) couplings. A pure state with no macroscopic order, the paramagnetic phase; a pure state with macroscopic order aligned with some mean bias in the couplings, the replica symmetric (RS) ferromagnetic phase (F); a macroscopically aligned phase, but with some complicated phase space fragmentation, the mixed phase (M); and a phase with no macroscopic alignment and a complicated fragmentation of the phase space, the spin glass phase (SG). Within both the sparse and dense Ising spin models these phases are exhibited and many features are shared by the two models.

The main question investigated in this paper is how phase behavior and transitions differ in the composite model from the sparse and dense frameworks, and whether a simple interpolation is produced by the composite models. Attention is restricted to cases in which the sparse sub-structure is percolating, since in any other regime the long range coupling will be due solely to the dense links. A non-percolating sparse substructure would not test the effects of competing long range induced order, although some of the methods and results are inclusive of this scenario and appear to vary continuously (at finite temperature) across the percolation threshold corresponding to the sparse substructure.

Section~\ref{composite.ensemble} outlines the ensemble of models studied, which are then analyzed by the replica method in section~\ref{composite.RM}. Section~\ref{composite.RS} develops the replica equilibrium solution by population dynamics alongside an analogous belief propagation (BP) algorithmic method applicable to typical samples. A stability analysis of the BP equations and population dynamics is derived. Section~\ref{composite.hightemp} presents a leading order solution to the composite system in terms of a simplified ansatz on the order parameter with results discussed in section~\ref{composite.leadingorderauxiliarysystems}. Section~\ref{composite.lowtemp} demonstrates the RS solutions for several composite models in the interesting range of parameters about the triple point in the phase diagram. Section~\ref{composite.finite} presents hypotheses on the structure of the low temperature phases alongside experimental 
results derived by BP and monte-carlo methods. We then conclude with a brief summary.

\section{\label{composite.ensemble} Composite ensembles}

The Composite model can be described by a Hamiltonian with coupling of $N$ spins
\begin{equation}
\Ham(\vS)= -\sum_{\ij} \left[J^D_{\ij} + J^S_{\ij}\right] S_i S_j - \sum_i \randomfield_i S_i\label{eq:composite.Ham}\;,
\end{equation}
where $\ij$ are an ordered set of variables. The couplings are labeled as dense ($D$) or sparse ($S$) and are sampled independently for each link according independent ensembles described shortly. The quenched variable abbreviation $\quenched$ indicates a sample of the couplings, and $\vS$ are the dynamic variables. The field vector $\vrandomfield$ is used only as a conjugate parameter to explore symmetries, the limit $\vrandomfield\rightarrow \vzeros$ (vector of zero fields) is always assumed throughout this chapter, although some physical quantities and insight are demonstrated using
conjugate fields as described in Appendix~\ref{app:ConjugateFields}.

The equilibrium properties of the model are studied. The Hamiltonian implies a static probability distribution on the state space given by
\begin{equation}
P(\vS) = \frac{1}{Z(\beta,\quenched)} \exp\left\lbrace -\beta
\Ham(\vS)\right\rbrace\label{eq:PS}\;,
\end{equation}
where $\beta$ is the inverse temperature and $Z$ is the partition function.

The spin states of interest are the typical case equilibrium distribution, in the large system limit. Properties of these states are established through the mean free energy
\begin{equation}
\beta \safed(\beta) = -\lim_{N \rightarrow \infty} \frac{1}{N}\< \log \partitionfunction \>_{\quenched} \label{eq:composite.safed}\;,
\end{equation}
where $\ensemble$ is the ensemble parameterization.

The model is fundamentally a fully connected one, the sparse component is realized as a subset of couplings that are an order of magnitude stronger. Due to this order of magnitude many results for standard densely connected spin models do not apply.

\subsubsection{\label{composite.SKSS} Dense (SK) sub-structure}

The dense sub-structure fully connects $N$ spin variables $\vS \in \left\lbrace \pm 1\right\rbrace^N$, with couplings sampled independently and at random according to the Gaussian distribution parameterized by $J_0$ and $J$
\begin{equation}
P(\mJ^D)= \prod_\ij P(J^D_\ij) \;; \qquad P(J^D_\ij) = \frac{1}{\sqrt{2\pi/N}} \exp \left\lbrace -\frac{N}{2 J^2}\left(J^D_\ij - \frac{J_0}{N} \right)\right\rbrace
\label{eq:composite.SK}\;,
\end{equation}
with a necessary scaling of components included. This set of couplings has a statistical description corresponding to the SK model.

\subsubsection{Sparse (VB) sub-structure}
\label{composite.VBSS}

It is convenient to factorize the sparse couplings as
\begin{equation}
 J^S_\ij= A_\ij \modulationsymbol_\ij\;.
\end{equation}
The ensemble is described by a connectivity matrix, $\mA$, which is zero for all but a fraction $C/N$ of components, and a dense coupling matrix $\mxi$, with no zero elements. In the irregular ensemble each directed edge is present (non-zero) independently with
probability $C/N$, with $C$ the mean variable connectivity, so that a prior for inclusion of an edge is
\begin{equation}
P(\mA)= \prod_\ij \left[\left(1-\frac{C}{N}\right)\delta(A_\ij) + \frac{C}{N}\delta(A_\ij -1)\right] \label{eq:composite.A}\;,
\end{equation}
this being the connectivity in a standard Erd\"{o}s-R\'{e}nyi random graph. The couplings in the non-zero cases are described by a distribution with finite moments, and are sampled independently according to
\begin{equation}
 P(\mxi)= \prod_\ij P(\modulationsymbol_\ij) \;; \qquad P(\modulationsymbol_\ij = x) = \phi(x) \label{eq:composite.phix}\;,
\end{equation}
in the general case. A practical distribution for analysis is the $\pm J$ distribution defined as
\begin{equation}
\phi(x)=(1-p)\delta(x-J^S) + p \delta(x+J^S) \label{eq:composite.pmJ}\;,
\end{equation}
with two parameters, $p$ the probability that the link is anti-ferromagnetic, and $J^S$ the strength of coupling. Regular connectivity ensembles have each variable constrained to interact with exactly $C$ neighbors,
\begin{equation}
 P(\mA) \propto \prod_{i=1}^N \delta\left(\sum_j A_{i j} - C\right) \label{eq:composite.HamAux4}\;.
\end{equation}

\subsubsection{Representative parameterizations}
\label{composite.specialcases}

Four models are considered in greater detail owing to their simplicity and ability to make transparent a range of observed phenomena. The F-AF model includes ferromagnetic dense couplings ($J=0$, $J_0>0$ (\ref{eq:composite.pmJ})) and anti-ferromagnetic sparse couplings ($p=1$ (\ref{eq:composite.pmJ})), with connectivity $C=2$, and is described by
\begin{equation}
\Ham(\vS)= - \frac{\mB(\gamma,J^S)}{N} \sum_\ij S_i S_j + J^S \sum_{\ij} A_\ij S_i S_j \label{eq:composite.HamAux1}\;.
\end{equation}
The function $\mB(\gamma,J^S)/N$ is introduced to balance the ferromagnetic and anti-ferromagnetic tendency. Choosing $\mB(\gamma,J^S)$ as a positive, monotonically increasing function of the scalar parameter $\gamma$ the relative strength of the anti-ferromagnetic and ferromagnetic parts are kept in some intuitive balance. As $\gamma$ increases there is an increased tendency towards aligning spins within the Hamiltonian -- the ferromagnetic (ordered) state is promoted.

It is also interesting to consider the converse case, the AF-F model with a ferromagnetic sparse part ($p=0$) and anti-ferromagnetic dense model ($J=0$,$J_0<0$), with connectivity $C=2$,
\begin{equation}
\Ham(\vS)= - J_S\sum_{\ij} A_\ij S_i S_j + \frac{\mB(\gamma,J^S)}{N} \sum_\ij S_i S_j \;, \label{eq:composite.HamAux2}
\end{equation}
with $\mB$ being again some suitably re-scaled function, $J^S$ must also be defined.

These models can also be considered for the case of regular connectivity. Either anti-ferromagnetic couplings, the {\em regular} F-AF (\ref{eq:composite.HamAux1}) and AF-F (\ref{eq:composite.HamAux2}) models, are considered, but in each case with connectivity chosen to be $C=3$ (a minimal choice above the percolation threshold).

\section{Replica method}
\label{composite.RM}

The replica method is used in both~\cite{Hase:DA,Raymond:OC} to study the composite system free energy in the limit of large $N$. The replica method is the most concise analytical method available, although many results presented herein can be developed through the cavity method with suitable assumptions. For convenience the fields
$\vrandomfield\rightarrow 0$, as in (\ref{eq:composite.Ham}), in the various calculation steps. Variations on this are useful in establishing a number of system properties as outlined in Appendix~\ref{app:ConjugateFields}.

In the replica approach the typical case behavior is examined through the free energy density (\ref{eq:composite.safed}) averaged over the quenched disorder. That is to say we do not expect typical samples from the ensembles to differ in the value of the order parameters and other extensive properties. The replica identity
\begin{equation}
\<\log \partitionfunction\>_{\quenched} = \left. \frac{\partial}{\partial n} \right|_{n=0} \repZ\;,
\end{equation}
allows for the average over the logarithm to be replace by the partition sum of a replicated set of variables. This is by an analytic continuation of $n$ to the set of integers, giving a form for which the quenched averages may be taken. The properties of the free energy are constructed through the replicated partition function
\begin{equation}
 \repZ = \prod_{\alpha=1}^n \left[\sum_{\vS^\alpha}\right]
\< \prod_{\ij} \exp \left\lbrace \beta(J^D_\ij +
 J^S_\ij) \sum_\alpha S_i^{\alpha} S_j^{\alpha}\right\rbrace\>_\quenched\;,
\end{equation}
where the quenched averages and dynamic averages may be taken equivalently.

The exponent is factorized with respect to the quenched variables in the sparse and dense parts. The average in the dense part involves an expansion to second order in $N$ of $J^D_\ij$. The leading order terms are described by $J_0$ and $J^2$ (\ref{eq:composite.SK}), and higher order terms are taken to be negligible in the large $N$ limit. The average in the sparse part is more involved, the full method is presented in Appendix~\ref{app:CompositeSystem_Replica}. The brief outline of the method in the remainder of this section applies only for Poissonian connectivity in the sparse sub-structure. The site dependence in the energetic part is factorized in general by introducing three classes of order parameters
\begin{equation}
\qal=\frac{1}{N}\sum_i S^\alpha_i \;; \qquad
\qalal=\frac{1}{N}\sum_i S^{\alpha_1}_i S^{\alpha_2}_i \;; \qquad \GENOP(\rvS) = \frac{1}{N}\sum_i \delta_{\rvS,\rvS_i} \label{eq:composite.GENOP}\;;
\end{equation}
where $\qal$ describes the homogeneous magnetization, $\qalal$ describes the 2-replica correlations, and the generalized order parameter~\cite{Monasson:OP} $\GENOP(\rvS)$ describes many kinds of spin correlations, where the bold font vector notation is used to represent a vector labeled by replica indices rather than site indices,
denoted by an over-line vector notation,
\begin{equation}
\delta_{\rvS,\rvS_i} = \prod_{\alpha=1}^n \delta_{S^\alpha,S^\alpha_i}\;;\qquad \rvS=\{S^\alpha | \alpha=1 \ldots n\}\;.
\end{equation}

The order parameters $\qal$ and $\qalal$ can be defined from the generalized order parameter in the Poissonian connectivity case
\begin{equation}
\qal = \sum_{\rvsigma} \GENOP(\vsigma) \sigma^\alpha \;; \qquad \qalal =\sum_{\rvsigma} \GENOP(\vsigma) \sigma^{\alpha_1} \sigma^{\alpha_2} \label{eq:composite.equi}\;.
\end{equation}
However, solving the saddle-point equations, by population dynamics in the RS
description, where order parameters are assumed to be invariant under replica-index permutations, is complicated without the redundant description (\ref{eq:composite.GENOP}), and the redundant description is necessary in the regular and F-F models. Furthermore having order parameters describing both dense and sparse parts is useful in discriminating effects due to sparse and dense sub-structures and the connection with the standard sparse and dense descriptions is also made transparent in the limiting cases: taking $\qal=\qalal=0$ to recover the thermodynamics of a sparse system; and $\GENOP(\rvsigma)=1$ to recover a purely dense thermodynamic description.

The original mixed topology problem is replaced by a site factorized (mean field) model - the complexity being encoded in a set of interactions amongst replica encoded in the order parameters. The definitions of the order parameters may be transformed to an exponential form by introducing a weighted integral over conjugate parameters (denoted by 
a hat). The exponential form allows a saddle-point method to be applied, an extremisation of the exponent allows the free energy to be identified as
\begin{equation}
\beta \safed = \lim_{n\rightarrow 0}\frac{\partial}{\partial n}\Extr_{\{\GENOP,\GENOPconj,\qal,\qhal,\qalal,\qhalal\}} \left\lbrace \Gone(\beta,\ensemble,\GENOP) + \Gtwo(\beta,\ensemble,\GENOPconj) + \Gthree(\GENOPconj,\GENOP) \right\rbrace \label{eq:composite.replicafreeenergy}\;,
\end{equation}
up to constant (ensemble parameter dependent) terms. The term $\Gone$ encodes an energetic term describing interactions, which in the absence of an external field is given by
\begin{equation}
\begin{array}{lcl}
 \Gone &=& - \frac{1}{2}\beta J_0 \sum_\alpha (\qal)^2 - \frac{1}{2}\beta^2 \! J^2 \sum_\alal (\qalal)^2 \\
&-& \frac{C}{2} \log \sum_{\rvS,\rvS'} \GENOP(\rvS)\GENOP(\rvS') \int \rmd x \phi(x) \exp\left\lbrace \beta x \sum_\alpha S^\alpha S'^\alpha \right\rbrace \label{eq:composite.G1}\;,
\end{array}
\end{equation}
where $\phi(x)$ is the coupling distribution in the sparse part (\ref{eq:composite.pmJ}). The term $\Gtwo$ is an entropic term coupling the sparse and dense order parameters
\begin{equation}
\Gtwo \!=\! - \log \sum_{\rvS}\exp\left\lbrace \!\sum_\alpha \qhal S_\alpha \!+\! \sum_{\alal} \qhalal S^{\alpha_1} S^{\alpha_2} \!+ C \GENOPconj (\rvS) \right\rbrace\;. \label{eq:composite.G2}
\end{equation}
The coupling between the order parameters and their conjugate forms is present in the term
\begin{equation}
\Gthree \!= \! C \sum_{\rvS} \GENOP(\rvS) \GENOPconj(\rvS) + \sum_\alpha\qal \qhal + \sum_{\alal} \qalal\qhalal \label{eq:composite.G3}\;.
\end{equation}

The free energy is used to calculate various self averaging properties of the system by taking derivatives with respect to conjugate parameter, as outlined in Appendix~\ref{app:ConjugateFields}. The inverse temperature is conjugate to the energy, from which the entropy is calculated. Derivatives with respect to uniform fields conjugate to $\vones$ can be used to test emergent ferromagnetic order. By inclusion of a random field of mean zero, the variance can be used to calculate correlation functions and susceptibility.

The order parameters, defined at the extrema of the saddle-point (denoted $*$), obey coupled saddle-point equations
\begin{equation}
\GENOP^*(\rvS)\! = \! \localreplicaprobability(\rvS); \qquad \qal^* = \sum_{\rvS} S^\alpha \localreplicaprobability(\rvS) ; \qquad \qalal^* =\sum_{\rvS} S^{\alpha_1} S^{\alpha_2} \localreplicaprobability(\rvS) \label{eq:composite.saddle1}\;,
\end{equation}
where
\begin{equation}
\localreplicaprobability(\rvsigma) \propto \exp\left\lbrace C \GENOPconj^*(\rvsigma)\! + \!\sum_\alpha \qhal^* \sigma^\alpha \!+\! \sum_\alal \qhalal^* \sigma^{\alpha_1} \sigma^{\alpha_2} \right\rbrace \label{eq:composite.saddlePrvsigma}\;,
\end{equation}
is a normalized probability distribution on the replicated state space.

The conjugate parameters are determined by equations without coupling between the sparse and dense parts
\begin{equation}
\GENOPconj^*(\vsigma) \propto \sum_{\rvtau} \GENOP^*(\rvtau) \< \exp \left\lbrace \beta x \sum_\alpha \tau^\alpha \sigma^\alpha \right\rbrace\>_x \;; \qquad \qhal^* = \beta J_0 \qal^* \;; \qquad \qhalal^* = \beta^2 J^2 \qalal^* \;; \label{eq:composite.saddle2}
\end{equation}
with $x$ distributed according to $\phi(x)$ as in (\ref{eq:composite.phix}). From these six equations it is possible to eliminate the conjugate parameters (\ref{eq:composite.saddle2}) to leave a fixed point defined without the conjugate parameters.

\section{Replica symmetric formulation and belief propagation}
\label{composite.RS}

\subsection{The RS saddle-point equations}
\label{composite.RSsaddle-point}

The order parameters are defined by the standard sparse and dense RS forms
\begin{equation}
\GENOP^*(\vsigma) = \int \rmd h \RSOP(h) \prod_{\alpha=1}^n
\frac{\exp\left\lbrace h \sigma^\alpha \right\rbrace}{2 \cosh
 h}\;;\qquad \qal^* = m \;; \qquad \qalal^* = q \;;
\end{equation}
with the variational aspects captured by the normalized distribution on the real line ($\RSOP$) and two scalar parameters ($m,q$).

The saddle-point equations can then be written for the general case, inclusive of regular and Poissonian connectivity , as
\begin{equation}
\RSOP(h) \propto \int \<\prod_{c=1}^{c_e} \left[\rmd h_c \rmd x_c \RSOP(h_c) \phi(x_c) \right] \delta\left(h - h^{RS} \right)\>_{c_e,\lambda}\label{eq:composite.RSsaddle1}\;,
\end{equation}
where
\begin{equation}
 h^{RS} = m + \lambda \sqrt{q} + \sum_{c=1}^{c_e} \atanh\left(\tanh(\beta x_c)\tanh(h_c) \right) \label{eq:composite.hRS}\;,
\end{equation}
and $c_e$ is distributed according to the excess connectivity distribution, a normalized distribution proportional to $C P(C-1)$, where $P(C)$ is the full variable connectivity distribution, regular or Poissonian. The integration variable $\lambda$ is normally distributed. The dense parts are defined similarly
\begin{equation}
m = \int \<\prod_{c=1}^{c_f} \left[\rmd h_c \rmd x_c \RSOP(h_c) \phi(x_c) \right] \delta\left(h - h^{RS}\right) \tanh(h)\>_{{c_f},\lambda}\label{eq:composite.RSsaddle2}\;,
\end{equation}
and
\begin{equation}
q = \int \<\prod_{c=1}^{c_f} \left[\rmd h_c \rmd x_c \RSOP(h_c) \phi(x_c) \right] \delta\left(h - h^{RS}\right)\tanh^2(h) \>_{{c_f},\lambda} \label{eq:composite.RSsaddle3}\;,
\end{equation}
but with the averages in ${c_f}$ being with respect to the
full connectivity distribution.

These equations can be solved by a method of population dynamics~\cite{Mezard:BLSG}
subject to two additional recursions on scalar quantities (\ref{eq:composite.RSsaddle2})-(\ref{eq:composite.RSsaddle3}).

\subsection{Composite belief propagation equations}
\label{composite.BP}

Composite BP can be interpreted in the context of the composite system as a heuristic method of determining marginals of the static probability distribution (\ref{eq:composite.P}), given a quenched sample~\cite{Kschischang:FG}. Whereas an exhaustive calculation requires $O(2^N)$ operations to construct a marginal, BP is guaranteed to produce an estimate in a number of operations that scales only linearly with the number of edges.

The equations from factors to nodes are trivial in the case of binary factors, so iterations on variable messages alone can be composed. Defining two directed messages for every link $\ij$, which can be interpreted as log-posterior ratios for spins on graphs with some interactions removed (cavity graphs)
\begin{equation}
h^{(t+1)}_{i \rightarrow j} =\frac{1}{2 \beta} \sum_{\tau_i} \tau_i \log {\hat P}^{(t+1)}(S_i=\tau_i | G_{i \rightarrow j}) = \frac{1}{\beta}\sum_{k \setminus \{i,j\}} \atanh\left(\tanh(\beta h^{(t)}_{k \rightarrow i})\tanh(\beta J_{\< i k \>})\right) \label{eq:composite.BP}\;,
\end{equation}
where ${\hat P}$ is used to denote an approximated probability distribution. The cavity graph is a factor graph rooted in variable $i$ with the coupling $J_\ij$ set to zero. The assumption underlying the probabilistic recursion is the independence of log-posterior ratios, which allows them to be used accurately as priors in each step, so that the recursion is equivalent to that on a tree.

BP can be iterated from some initial condition. If correlations between messages are sufficiently weak then the messages will converge to correctly describe the probabilities. From these marginal properties, such as the magnetization at equilibrium, can be constructed. A log-marginal may be estimated by
\begin{equation}
H^{(t+1)}_i = \frac{1}{2\beta} \sum_\tau \tau \log {\hat P}^{(t+1)}(S_j=\tau | G) = \frac{1}{\beta}\sum_{j \setminus i} \atanh\left(\tanh(\beta h^{(t)}_{j \rightarrow i})\tanh(\beta J_\ij)\right) \label{eq:composite.H}\;.
\end{equation}

The condition of sufficiently weak correlations is closely related to the notion of a pure state is statistical mechanics~\cite{Mezard:SGT}. The assumption of independent messages applies only when the log-posteriors (\ref{eq:composite.BP}) reflect the distribution in a pure state, the similarity with (\ref{eq:composite.hRS}) is not coincidental. Pure states act as local attractors of the BP dynamics, and it is only when there is a competition between these attractors that dynamics is expected to fail. With BP initialized sufficiently close (globally) to a pure state, or in the case of a unique attractor, convergence to the pure state can be anticipated giving a correct description of the equilibrium probability distribution.

\subsubsection{Simplification of dense messages}
\label{composite.SDM}
\begin{figure*}
\begin{center}
\includegraphics[width=\linewidth]{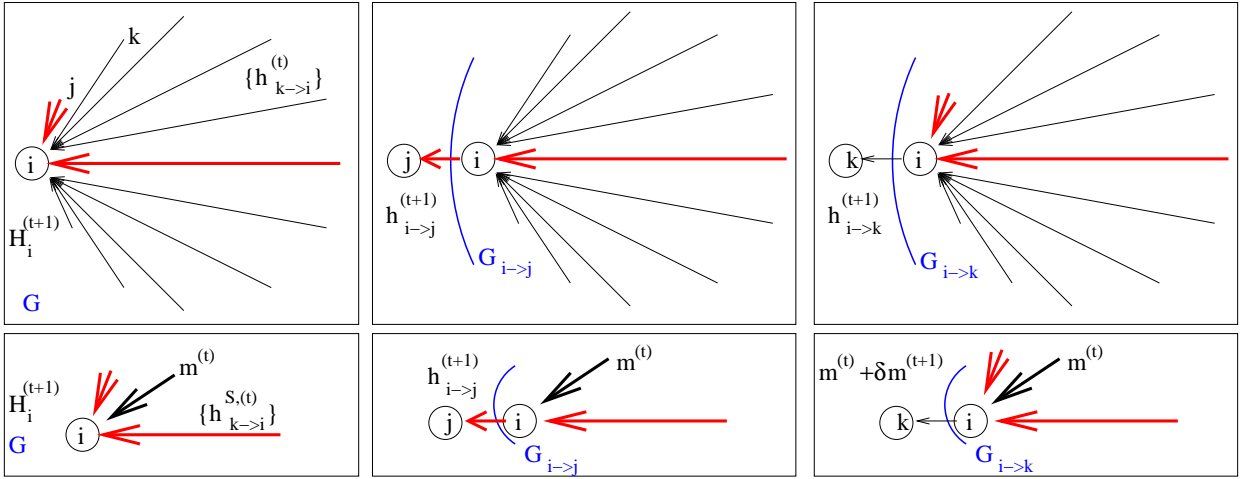}
\caption[Composite BP]{(color online) \label{fig:composite.BP} BP constructs an estimate of the posteriors by message passing, each message is a log-posterior estimate for some variable subject to the removal of the interaction with another variable (the variable to which the message is passed), as in the top sub-figures. In the lower two sub-figures the central limit is applied to the messages on dense links and in some cases only a single parameter is then required to represent the $O(N)$ dense messages. A related approximation is implicit in the derivation of the RS free energy.}
\end{center}
\end{figure*}

Assuming the messages to be independent, then each message can be considered as a random object determined by the couplings in the cavity graph. The messages are therefore i.i.d. and the sum over many messages will converge to a Gaussian random variable. To leading order the messages may be rewritten incorporating this insight
\begin{equation}
h^{(t+1)}_{i \rightarrow j} = m^{(t)} + \sqrt{q^{(t)}}\lambda^{(t)}_{i \rightarrow j} + \frac{1}{\beta}\sum_{k \in \{ \partial_i \setminus j\}} \atanh\left( \tanh(\beta h^{(t)}_{k \rightarrow i}) \tanh(\beta J_\ij)\right) \label{eq:composite.gaussian}\;,
\end{equation}
where $m^{(t)}$ is the mean and $q^{(t)}$ the variance, and term $\partial_i$ is used to denote variables connected to $i$ through strong couplings. The distribution over
reweighted messages $\lambda_{i\rightarrow j}$ will be asymptotically Gaussian if the approximation is correct. The value of the message for a particular instance of the quenched disorder is given by
\begin{equation}
m^{(t)} + q^{(t)}\lambda^{(t)}_{i \rightarrow j} = \frac{1}{\beta}\sum_{k \setminus \{ \partial_i \cup j\}} \atanh\left( \tanh(\beta h^{(t)}_{k \rightarrow i}) \tanh(\beta J_\ij)\right)\label{eq:composite.lambda_ij}\;.
\end{equation}
The Gaussian statistics are defined by analogy with the RS thermodynamic quantities, to leading order in $N$
\begin{equation}
m^{(t)} = \beta J_0 \frac{1}{N}\sum_{i=1}^N \tanh(\beta
 H_i^{(t)}) \;; \qquad q^{(t)} = \beta^2 J^2
\frac{1}{N}\sum_{i=1}^N \tanh^2(\beta H_{i}^{(t)})\;,
\end{equation}
for any dense set of couplings~\cite{Kabashima:PB}. The log-posterior ratios for the spin states on the full graph are approximated as
\begin{equation}
\beta H^{(t+1)}_{j} = m^{(t)} + \sqrt{q^{(t)}}\lambda^{(t)}_{j} + \sum_{k \in \partial_i} \atanh\left( \tanh(\beta h^{(t)}_{k \rightarrow i})\tanh(\beta J_\ij) \right) \label{eq:composite.Hgaussian}\;.
\end{equation}
The term $\lambda_i$ is closely related to $\lambda_{i \rightarrow j}$, up to a correction of order $1/N$, by removing the restriction on the sum in $j$ from (\ref{eq:composite.lambda_ij}).

In the case that $J\neq 0$ it is necessary to evaluate $\lambda_{i}$ for each link, still requiring $O(N^2)$ evaluations as in the original algorithm. To reduce computational complexity it may be valuable to marginalize over this if $J \ll J_0$ or if the sparse couplings dominate dynamics, but if $J=0$ it is sufficient to take $\lambda_{i}^{(t)}=0$ and algorithm complexity is reduced to $O(N)$, as illustrated in figure~\ref{fig:composite.BP}. A method for combining messages in models comprising both densely and sparsely interacting components has been recently introduced~\cite{Mallard:BPDG}.

\subsection{Stability analysis}
\label{composite.Stability}

If the replica description correctly describes a single pure state, then this implies the spin glass susceptibility is not divergent in the thermodynamic limit. In the case of a sparse graph the tree-like approximation provides a natural basis for constructing a self-consistent estimate of the spin-glass susceptibility~\cite{Rivoire:GM}, whereas in the dense model a direct test of eigenvalue stability towards replica symmetry breaking can establish a complete description~\cite{Almeida:SSK}.

An analytic framework entirely within the replica method might be constructed to test
spin-glass susceptibility. As in Appendix~\ref{app:ConjugateFields}, a connection can be made between the particular instability in the order parameter and the divergence of the physical quantity, spin glass stability, within the RS framework. This identity is not pursued within this paper, instead a more intuitive framework, believed to be equivalent, is presented.

The non-divergence of the spin-glass susceptibility in sparse and dense models requires the local stability of the saddle-point equations; this proves to be an equivalent condition to the stability of the BP equations on a typical graph in the limit $N\rightarrow\infty$~\cite{Rivoire:GM,Kabashima:PB}. Stability of the BP equations is therefore explored for a typical sample. Assuming a linear perturbation $\{\delta h^{(t)}_{i\rightarrow j}\}$ about some fixed point $\{ h^{(t)}_{i\rightarrow j}\}$ of the BP equations (\ref{eq:composite.BP}), implies an independent recursion on the perturbations that may be written at leading order
\begin{equation}
 \delta h^{(t+1)}_{j\rightarrow k} = \sum_{i \setminus \{j,k\}} \delta h^{(t)}_{i\rightarrow j} \frac{(1-\tanh^{2}(\beta h^{(t)}_{i \rightarrow j}))\tanh(\beta J_\ij)}{1-\tanh^2(\beta h^{(t)}_{i\rightarrow j}) \tanh^2(\beta J_\ij)}\label{eq:composite.BPstab}\;.
\end{equation}
In the dense part, the fluctuations may again be represented by a Gaussian random variable of mean and variance
\begin{equation}
J_0 \<\delta h_{i\rightarrow j}^{(t)} (1-\tanh^2(\beta h_{i\rightarrow j}^{(t)}))\>\; \mbox{ and } \qquad \<(\delta h_{i\rightarrow j}^{(t)})^2 (1-\tanh^2(\beta h_{i\rightarrow j}^{(t)}))^2\>\;,
\end{equation}
respectively, since the couplings are assumed to be uncorrelated with the perturbations in BP, the average is with respect to all perturbations and fields incident on $j$. An expansion of $h_{i \rightarrow j}$ in terms of $H_i$ is possible so that the statistics can be shown to be identical at leading order for all $j$~\cite{Kabashima:PB}, therefore the perturbations evolve according to quantities which are time but not site dependent
\begin{equation}
\delta m^{(t)} = J_0 \<\delta H_i^{(t)}\left( 1 - \tanh^2(\beta H_i^{(t)})\right) \>\;; \qquad \delta q^{(t)} = J^2 \<(\delta H_i^{(t)})^2 \left(1-\tanh^2(\beta H_i^{(t)})\right)^2 \> \;;\label{eq:composite.perturbations}
\end{equation}
where $\delta H_i^{(t)}$ are the perturbations in the log-posteriors, which are equal to $\delta h^{(t)}_{i \rightarrow j}$ at leading order whenever $J_\ij$ is not a strong coupling term.

A final approximation is to assume $H_i$ is uncorrelated with $\delta H_i$. In this case the statistics can be written only in terms of $q^{(t)}$, $\<\delta H_i\>$ and $\<(\delta H_i)^2\>$. However, this is not true at leading order when a sparse component is present. Variables with larger connectivity in the sparse part, are described by a field
distribution of greater variance, and the perturbations scale similarly. Instead, the pair of correlation functions (\ref{eq:composite.perturbations}) determines the evolution of perturbations.

Evolution of the perturbations can be undertaken in parallel with BP; to each message is attached a representative statistic for, or a distribution over, perturbations. It is sufficient to consider a distribution of perturbations characterized by a mean $\bar{\delta h}_{i \rightarrow j}^{(t)}$, and variance $\bar{\delta h^2}_{i \rightarrow j}^{(t)}$, attached to each macroscopic field. If these parameters decay exponentially, in expectation, then this is an indication of fixed point stability.

Assuming that there is no linear instability, the equation determining $\bar{\delta h^2}_{i \rightarrow j}^{(t)}$ is
\begin{equation}
\bar{\delta h^2}_{i \rightarrow j}^{(t+1)} = \delta q^{(t)} + \sum_{i \in \partial_j \setminus k} \bar{\delta h^2}_{i \rightarrow j}^{(t)} \left(\frac{(1-\tanh^{2}(\beta h^{(t)}_{i \rightarrow j})) \tanh(\beta J_\ij)}{1-\tanh^2(\beta h^{(t)}_{i\rightarrow j}) \tanh^2(\beta J_\ij)} \right)^2 \label{eq:composite.VarianceProp}\;,
\end{equation}
with a similar equation applicable to the case of a linear perturbation.

The BP equations can be interpreted as a recursive instantiation of the RS saddle-point equations (\ref{eq:composite.RSsaddle1})-(\ref{eq:composite.RSsaddle3}) except in the explicit site dependence, so that quenched disorder specific correlations may accumulate over several updates. Assuming a negligible feedback process in BP, or a modified problem without loops or with annealed disorder, the macroscopic properties established by BP will depend only on the steady state distribution of messages on sparse links and the mean and variance of dense messages. Objects analogous to a histogram estimate to $\RSOP$ (\ref{eq:composite.RSsaddle1}), and scalar parameters $m^{(t)}$ and $q^{(t)}$ in the saddle-point method. However, at the level of the mapping of individual points in the RS description (\ref{eq:composite.hRS}) it is possible that local fluctuations of the messages on fields are unstable, despite stability in the distribution. Whereas divergence in $\<\bar{\delta h}\>$ might be observed in a macroscopic instability in the first moment of $\RSOP$, an instability of the mapping in $\<\bar{\delta h^2}\>$ will not be realized in any macroscopic moment of the distribution. It is this instability in the mapping which is probed by the BP stability analysis. In the absence of a linear instability it is assumed divergence in $\<\bar{\delta h^2}\>$ is a necessary condition for any local instability.

The fluctuations on sparse messages are represented fully in this framework, whereas dense messages are summarized under approximation. The stability is a self-consistent (longitudinal) test of stability, but is known not to probe all possible instabilities and so provides only a sufficient criteria for instability. The SK model is an example where the longitudinal stability of the ferromagnetic phase, as derived through a BP framework~\cite{Kabashima:PB}, does not capture correctly the spin glass transition at low temperature, as shown in figure~\ref{fig:composite.StandardDiag}. Since the models investigated in detail later have inhomogeneity in the sparse sub-structure only ($J^2=0$), it is felt the test of stability as applied in this paper may be a more accurate reflection of true local stability towards replica symmetry breaking. A connection between the stability tested through the BP framework, and an instability entirely within the replica method, might be established as outlined in Appendix~\ref{app:ConjugateFields}.

\section{Exact high temperature formulation}
\label{composite.hightemp}

In the limit $\beta\rightarrow 0$ the paramagnetic solution $\GENOP=1,\qal=0,\qalal=0$ is the only stable solution, but becomes unstable as temperature is decreased. This process can be investigated by considering the moments of $\GENOP$ through a moment expansion representation
\begin{equation}
\GENOP(\rvsigma) = 1 + \sum_\alpha \qbal \sigma^\alpha + \sum_\alal \qbalal \sigma^{\alpha_1} \sigma^{\alpha_2} + \sum_{L=3} \sum_\orderedL{\alpha}  {\bar q}_\orderedL{\alpha} \sigma^{\alpha_1}\ldots \sigma^{\alpha_L} \label{eq:composite.GENOPexpanded}\;.
\end{equation}
The saddle-point equations can be solved in each moment $\left\lbrace {\bar q}\right\rbrace$, and stability tested in some subset of the moments.

In the sparse sub-structure both the excess and full connectivity distributions are Poissonian, the saddle-point equation (\ref{eq:composite.saddle1}) can be expanded, using the identity (\ref{eq:composite.equi}), as
\begin{equation}
P(\rvsigma) = \prod_{L=1}^\infty \left[\prod_{\orderedL{\alpha}} \left[\cosh( X_L {\bar q}_{\orderedL{\alpha}}) (1 + \sigma^{\alpha_1} \cdots \sigma^{\alpha_L} \tanh(X_L {\bar q}_{\orderedL{\alpha}}) \right]\right] \label{eq:composite.P}\;,
\end{equation}
eliminating the conjugate parameters (\ref{eq:composite.saddle2}). The terms
\begin{equation}
X_1 = \beta J_0 + T_1\;; \qquad X_2 = \beta^2 J^2 + T_2\;; \qquad X_i = T_i\; \hbox{ if } \; i>2 \label{eq:composite.X}\;,
\end{equation}
determine transition properties where
\begin{equation}
T_i = C \int \rmd x \phi(x) \tanh^i(\beta x)\;.
\end{equation}
The saddle-point equations can be written in terms of an equation for each moment
\begin{equation}
\bq_\orderedL{\alpha} \! = \! \tanh(X_L \bq_{\orderedL{\alpha}}) \!+ \! \frac{(1-\tanh^2(X_L \bq_{\orderedL{\alpha}}))\< S^{\alpha_1} \ldots S^{\alpha_L}\>_{\sim \bq_{\orderedL{\alpha}}} X_L}{1 \! + \< S^{\alpha_1} \ldots S^{\alpha_L}\>_{\sim \bq_{\orderedL{\alpha}}} \tanh(X_L \bq_{\orderedL{\alpha}})}\;,\label{eq:composite.saddleexpanded}
\end{equation}
where the notation $\<\cdots\>_{\sim x}$ indicates an average with respect to (\ref{eq:composite.P}), but with $x=0$. A solution is apparent which is the paramagnetic solution with $z=\<\sigma^{\alpha_1} \ldots \sigma^{\alpha_L}\>$ and
 $q_\orderedL{\alpha} =0$ for all choices of indices. This is the only solution when
 $X_L\rightarrow 0$, corresponding to the high temperature limit.

At lower temperature a solution may emerge in one of the moments. It is only necessary to show that some component $\bq$ allows a non-zero solution. The second term in (\ref{eq:composite.saddleexpanded}) is zero at leading order in $\bq$ in the moments of
the distribution, and there is no coupling of the moments at leading order. Hence any solution which emerges continuously from the paramagnetic solution must do so with equality at leading order between the first term of the right hand side and the left hand side. This leads to a criteria $X_L=1$ for the existence of a continuous transition.

For a discontinuous transition to occur in some component, without $X_i > 1$, requires the derivative of the second part with respect to $\bq$ to be a convex function of $\bq$ in some range of the parameter (\ref{eq:composite.saddleexpanded}). However, the
derivative is a concave function of $\bq$, so that unless $X_i>1$ for some component, there can be no solution other than the paramagnetic one.

\subsection{High temperature phase transitions}
\label{composite.HTPT}

The existence of non-paramagnetic order is determined from (\ref{eq:composite.saddleexpanded}) as:
\begin{equation}
\begin{array}{lcl}
 X_1 &>& 1 \qquad\hbox{1-spin / Ferromagnetic (F) order}\;;\\
 X_2 &>& 1 \qquad\hbox{2-spin / Spin Glass (SG) order}\;;\\
 X_L &>& 1 \qquad\hbox{L-spin order}\;.
\end{array}\label{eq:composite.HTT}
\end{equation}
In each case the solution which emerges may be estimated by an expansion in the right hand side of (\ref{eq:composite.saddleexpanded}) up to some order. Cubic order can be considered as a minimum to obtain the continuously emerging solution.
To allow for the replica limit $n \rightarrow 0$ an assumption on the correlations is required, RS being the simplest, and the order parameters may then be determined. Depending on the order of solution required, some coupling between moments is relevant, and it is necessary to solve a set of coupled equations.

The emergence of a ferromagnetic phase is realized in a continuous transition towards non-zero values of $\qbal$. Through coupling of the order parameters, all parameters $\bq_\orderedL{\alpha}$ become non-zero at order $O((\qbal)^L)$.

The emergence of a spin glass phase is realized in a continuous transition towards non-zero values of $\qbalal$, while $\qbal=0$. Even order parameters are generated by the coupling of terms at higher order.

The transition towards an $L$-spin order is irrelevant to the high temperature analysis, since by consideration of (\ref{eq:composite.X}) it is clear that $X_L \leq X_2$ for all $L>2$, with equality only in pathological cases, therefore the transition can only be towards a ferromagnetic or spin glass phase.

In the case that $X_1=X_2$, at the high temperature transition point both orders may emerge simultaneously and in competition. This case can be understood at leading order through an SK auxiliary model.

\subsection{SK auxiliary system}
\label{composite.auxiliary}
\begin{figure*}
\centering
 \includegraphics[width=\linewidth]{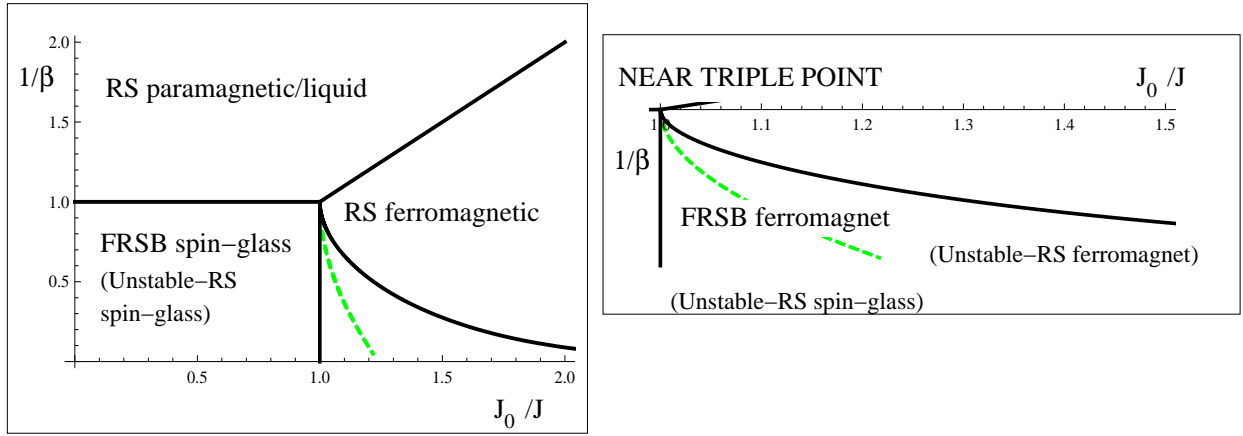}
\caption[SK phase diagram.]{(color online) \label{fig:composite.StandardDiag} The phase diagrams for disordered spin glass systems often exhibit a phase behavior similar to the SK model. Left figure: The phase transitions are indicated by solid dark lines. As temperature is lowered there is a transition from an RS paramagnetic phase ($m=q=0$) to either an RS ferromagnetic ($m>0$) or spin glass ($q>0$,$m=0$) phase. As temperature is lowered in the ferromagnetic phase there is also an RS to Full-RSB transition. Under the RS assumption the longitudinal instability measures calculated in the context of BP coincides with the F-SG transition in the RS description (dashed line). The instability of the ferromagnetic 
phase is not correctly predicted; the result is a lower bound in temperature for the replica instability in the ferromagnetic phase (towards a mixed phase).}
\end{figure*}

In either the case of a ferromagnetic order, or spin glass order, the behavior is described at leading order about the paramagnetic phase by the terms $\left\lbrace \qal\right\rbrace$ and $\left\lbrace \qalal\right\rbrace$. The free energy can be written in these cases as a function of only these two types of order parameter. After elimination of conjugate parameters the free energy can be written up to constant terms as
\begin{equation}
\beta \safed \!=\! \lim_{n \rightarrow 0} \frac{\partial}{\partial n} \!\left(\!-\! \log \!\!\sum_{\rvS} \! \exp \!\left\lbrace \! X_1 \!\sum_\alpha q_\alpha S^\alpha \! + \! X_2\!\!\! \sum_\alal \qalal S^{\alpha_1} S^{\alpha_2}\!\!\right\rbrace \!+\! \frac{X_1}{2}\!\! \sum_\alpha q_\alpha^2 \!+\! \frac{X_2}{2}\!\! \sum_\alal \!\! \qalal^2 \!\right) \label{eq:composite.SKauxiliary}\;.
\end{equation}
This is the replica formulation of the SK model free energy~\cite{Sherrington:SMSG}. Therefore at leading order the high temperature phases are equivalent to the SK model, up to the $\beta$ dependence of the energetic coupling terms. Instead of the standard term $\beta J_0$ there is $X_1$, and instead of $\beta^2 J^2$ there is $X_2$.

For every composite system of Poissonian connectivity there exists an auxiliary SK model with an equivalent leading order behavior at high temperature. By mapping the composite parameterizations to the SK model all the leading order high temperature transition properties must carry over, including the nature of RSB and the stability of the RS description.

Let $A$ denote the parameterizations $(J^A_0,J^A,\beta^A)$ of an SK model with an equivalent high temperature behavior to some composite system at the high temperature transition. This parameterization is redundant, there are only two independent parameters and so $J^A=1$ is chosen. The standard phase diagram for an SK model under these parameterization is demonstrated in figure~\ref{fig:composite.StandardDiag}.

The auxiliary parameterization is determined by the mapping equilibrating the coefficients in the free energy (\ref{eq:composite.X})
\begin{equation}
\beta^A J^A_0 = X_1 \;; \qquad (\beta^A)^2 = X_2
\label{eq:composite.AuxiliaryMapping}\;.
\end{equation}
Where this mapping is continuous it is possible to consider how the auxiliary system parameterization responds to variation of temperature (or some other parameter) in the composite system.
Variation of $\beta$ in the composite model is realized as a trajectory in the auxiliary model parameter space given by
\begin{equation}
\frac{\partial J_0^A}{\partial \beta^A} = 2\frac{J_0 - J_S C
(1-\tanh^2(\beta J^S)}{J^S C \tanh(\beta J^S)(1-\tanh^2(\beta
J^S))} - \frac{1}{\beta^A} \label{eq:composite.partialJ0ApartialbetaA}\;.
\end{equation}

In the case that the couplings to higher order moments are small ($X_L \ll 1$ for $L>2$), then the mapping may be applied with some confidence to lower temperature. Such a scenario will occur when the $X_1$ and $X_2$ are dominated by the dense sub-structure terms, or when $C$ is large in the sparse sub-structure.

\subsection{Beyond leading order}
\label{composite.BLO}

The leading order approximation to the composite system differs from the SK model in the anomalous dependence of energetic components on $\beta$. This observation alone is sufficient to account for many of the novel features of composite models reported at high temperature.

About the ferromagnetic transition the term $\qal$ appears at leading order to provide a thermodynamic description. The magnitude of $(\qal)^2$ is proportional to $\Delta_1=X_1-1$ at leading order and at $L^{th}$ order the value is dependent on moments of the distribution up to $\bq_{\orderedL{\alpha}}$. The set of non-linear coupled equations can be solved in parallel at each order. The ferromagnetic phase is at leading 
order an RS phase so that an expansion with simple RS components will be stable at leading order. The full description of the ferromagnetic phase differs from the auxiliary system description at third or fourth order.

The spin glass phase does not include any non-zero odd moments, and is described at leading order by $\Delta_2=X_2-1$, and at second order includes the term $\qbalalalal$. This term arises from the sparse sub-structure and so behavior deviates from the auxiliary model at second order. However, since even moments have positive coefficients, all with a monotonic dependence on $\beta$, phenomenological properties may not differ
significantly from the VB model which has been thoroughly studied (e.g.~\cite{Mottishaw:RSB}).

In the vicinity of the triple point, where both $\Delta_1$ and $\Delta_2$ are positive, the terms $\qbalalal$ and $\qbalalalal$ are relevant at second order. The literature developed in studying the VB model is sufficient to describe RS properties, and stability about the triple point~\cite{Viana:PD,Mottishaw:SRF}. The leading order behavior gives a transition from an RS ferromagnet to a spin glass according to a balance in the components $\Delta_1= \Delta_2/2$. The second order term in the sparse model indicates the existence of a mixed phase, with a refinement of the transition line.

The AT line is sufficient to describe stability of an RS solution in the dense model at all temperatures~\cite{Almeida:SSK}. In order to correctly describe transitions in the sparse or composite models it is necessary to consider a wider range of eigenvalues~\cite{Mottishaw:SRF}, which cannot be evaluated other than numerically, except at the percolation threshold (absent in the composite model) or as a polynomial expansion truncated at some order.

A stability analysis considering moments up to fourth order was recently presented~\cite{Raymond:OC}. It considers an RS description with inclusion of second order effects $\{\qbalalal,\qbalalalal\}\neq 0\}$, but with an analysis of instabilities restricted to variation in $\{\qbal,\qbalal\}$. This predicts a comparable splitting of the line $\Delta_1=\Delta_2/2$ to those found for the VB model, but for some ranges of parameters a stable spin glass phase is incorrectly identified. Since only a restricted set of eigenvalues is considered this is not unreasonable, but demonstrates a weakness in the method.

\subsubsection{Regular connectivity}

The derivations of this section so far, beginning from
(\ref{eq:composite.saddleexpanded}) onwards have been specific to the case of Poissonian connectivity (\ref{eq:composite.G2}) and do not necessarily extend to composite systems with non-Poissonian connectivity. The replica theory is developed along similar lines to previous sections in Appendix~\ref{app:CompositeSystem_Replica} to be inclusive of the regular connectivity ensemble. The 1-spin and 2-spin dense sub-structure order parameters are determined by (\ref{eq:composite.GENOP}) and take zero values in the paramagnetic phase. The sparse sub-structure order parameter is different from (\ref{eq:composite.GENOP}) to be inclusive of non-Poissonian connectivity, but in general 
takes a value $\GENOP=1$ in the paramagnetic solution, and may be expanded as a set of moments (\ref{eq:composite.GENOPexpanded}). However, with the new definition $\qal\neq\qbal$ and $\qalal\neq\qbalal$ in general. Each of these order parameters
corresponds to distinct physical quantities: $\qal (\qalal)$ are related to the mean magnetization (2-spin correlation), whereas $\qbal,\qbalal$ correspond to these quantities weighted by connectivity in the sparse sub-structure, as indicated in Appendix~\ref{app:ConjugateFields}.

Along similar lines to the previous analysis, it is possible to consider the emergence of order by treatment only of the leading order behavior about the paramagnetic solution.
The 1-spin order terms are coupled at leading order by the saddle-point equations, thus there is no decoupled representation describing emergence of spin glass and ferromagnetic order in general. The criteria for a ferromagnetic solution to emerge continuously from the paramagnetic solution as temperature is lowered is determined by the point at which
\begin{equation}
\left(\begin{array}{c} \qal \\ \qbal \end{array}\right) = \left(\begin{array}{cc} \beta J_0 & T_1 \tanh(\beta x)\\
\beta J_0 & \frac{(C-1)}{C} T_1 \end{array} \right) \left(\begin{array}{c} \qal \\ \qbal \end{array}\right) \label{eq:composite.stability}\;;
\end{equation}
if such a point exists; its existence requires the principal eigenvector of the matrix to be the one vector. However, the existence of a solution point in the coupled equations is not guaranteed, and there exist a range of parameters in which decreasing temperature results in a pair of complex conjugate eigenvalues which exceed one in modulus.

The right hand side of (\ref{eq:composite.stability}) represents the leading order 1-spin terms in the saddle-point equations (\ref{eq:composite.saddle1}), after elimination of the conjugate parameters. In the case of Poissonian connectivity the existence of a continuous transition is necessary for the existence of a ferromagnetic or spin glass phase (\ref{eq:composite.saddleexpanded}). This is due to the concavity of the saddle-point equation when interpreted as a mapping, concavity is assumed to hold also for the regular connectivity composite system.


However, in a general model it is necessary only for the principal eigenvalue of the matrix~(\ref{eq:composite.stability}) to exceed one for some non-paramagnetic solution of the saddle-point equations to exist. 
When the modulus of the principal eigenvalue exceeds one the assumption of weak coupling between the moments in the order parameters expansion ceases to be valid when considering an expansion coincident with the eigenvector. The criteria that the modulus in the leading order expansion is greater than one corresponds to a set of criteria
\begin{equation}
\begin{array}{lcllr}
\frac{1}{2}\left| \left(\beta J_0 + \frac{C-1}{C}T_1\right) \pm \sqrt{\left(\beta J_0 + \frac{C-1}{C}T_1\right)^2 + 4 \frac{\beta J_0 T_1}{C}} \right| &>& 1 &&\hbox{1-spin order}\;; \\
\frac{1}{2}\left| \left(\beta^2 J^2 + \frac{C-1}{C}T_2\right) \pm \sqrt{\left(\beta^2 J^2 + \frac{C-1}{C}T_2\right)^2 + 4 \frac{\beta^2 J^2 T_2}{C}} \right| &>& 1 &&\hbox{2-spin order}\;; \\
\frac{C-1}{C} T_L &>& 1 &&\hbox{L-spin order}\;.
\end{array} \label{eq:composite.HTT3}
\end{equation}
The potential exists for the modulus to exceed one whilst the discriminant is less than
zero in the 1-spin order term, when either $T_1$ or $\beta J_0$ are negative. This phenomena, absent in the VB and SK models, is contingent on a sub-set of couplings being anti-ferromagnetic. In spite of a comparable functional form in the 2-spin order term, the transition from a paramagnet to a spin glass is always described by a non-negative discriminant, and real eigenvalues.

The complex eigenvalues imply complex conjugate eigenvectors. Where the eigenvalues are real it is possible to test the stability of the equilibrium solution by inclusion of a conjugate field in proportion to the eigenvector components (see Appendix \ref{app:ConjugateFields}).
However, where the eigenvalue is complex such a field is not physical and is not consistent with assumptions made in the development of the equilibrium solution.

Attention is restricted to real valued perturbations of the order parameters, which can be associated with the real valued conjugate fields. A local instability in the paramagnetic solution is only anticipated towards a ferromagnetic phase when the real part of the principal eigenvalue is larger than one, or towards a spin-glass solution when criteria (\ref{eq:composite.HTT3}) is met. If the paramagnetic solution is stable with respect to an infinitesimal term conjugate to the magnetization in the Hamiltonian then the paramagnetic solution will be recovered continuously as the conjugate field approaches zero. This is equivalent to the criteria that the linearized saddle-point equations are convergent to the zero solution.  Linear instability is apparent when the real-part of the eigenvalues exceed one. However, since the perturbation is not coincident with an eigenvector there is no leading order solution to the linearized equations when the external field is added. The instability in the paramagnetic solution is towards a discontinuously emerging solution.

The discontinuously emerging solution from the paramagnetic instability might be a locally stable (thermodynamic or metastable) solution across a wider range of temperatures than that indicated by the local stability analysis of the paramagnetic solution. In limited simulations, comparable in size to those described in section~\ref{composite.finite}, the behavior observed at temperatures close to (but below) the modulus one criteria (\ref{eq:composite.HTT3}) is consistent with the hypothesis of two locally stable solutions.
One solution describes the thermodynamic phase, and the other a metastable solution, with decreasing temperature a discontinuous thermodynamic is anticipated.

%

The case of large $\gamma$ allows only for a transition from a paramagnetic to ferromagnetic state, and this may be discontinuous. As well as a thermodynamic solution, several dynamical transitions may describe changes in local stability criteria of the solutions; these local instabilities may dominant aspects of dynamics, and in general will not be coincident with thermodynamic transitions.

At intermediate $\gamma$ values the paramagnetic solution may be locally unstable first towards a spin glass solution as temperature is lowered. The presence of another metastable or thermodynamic ferromagnetic phase may change the properties of this transition by comparison with the standard continuous case.

In the limit of large $C$, a simplified description is possible in the transition criteria for the regular connectivity case. With a sensible scaling of the moments of $\phi(x)$ so that $T_1$ and $T_2$ remain finite as $C$ increases, the final term in the discriminant (\ref{eq:composite.HTT3}) becomes negligible and a simple transition criteria is recovered, consistent with the Poissonian system
\begin{equation}
 \beta J_0 + T_1 > 1\;; \qquad
 \beta^2 J^2 + T_2 > 1 \;. \label{eq:composite.reglimHTT}
\end{equation}
This is also the result that would be obtained in naively applying the dense system method, using only a mean and variance of link strengths, to the two scale system. Examples of discontinuous high temperature transitions are examined in
section~\ref{composite.DHTT}, with a clear departure from the conditions laid out in (\ref{eq:composite.reglimHTT}).

\section{Leading order predictions for phase behavior}
\label{composite.leadingorderauxiliarysystems}

\subsection{The F-AF model}
\label{composite.FAFauxiliarysystem}
\begin{figure*}[!htbp]
\centering{
 \includegraphics[width=0.8\linewidth]{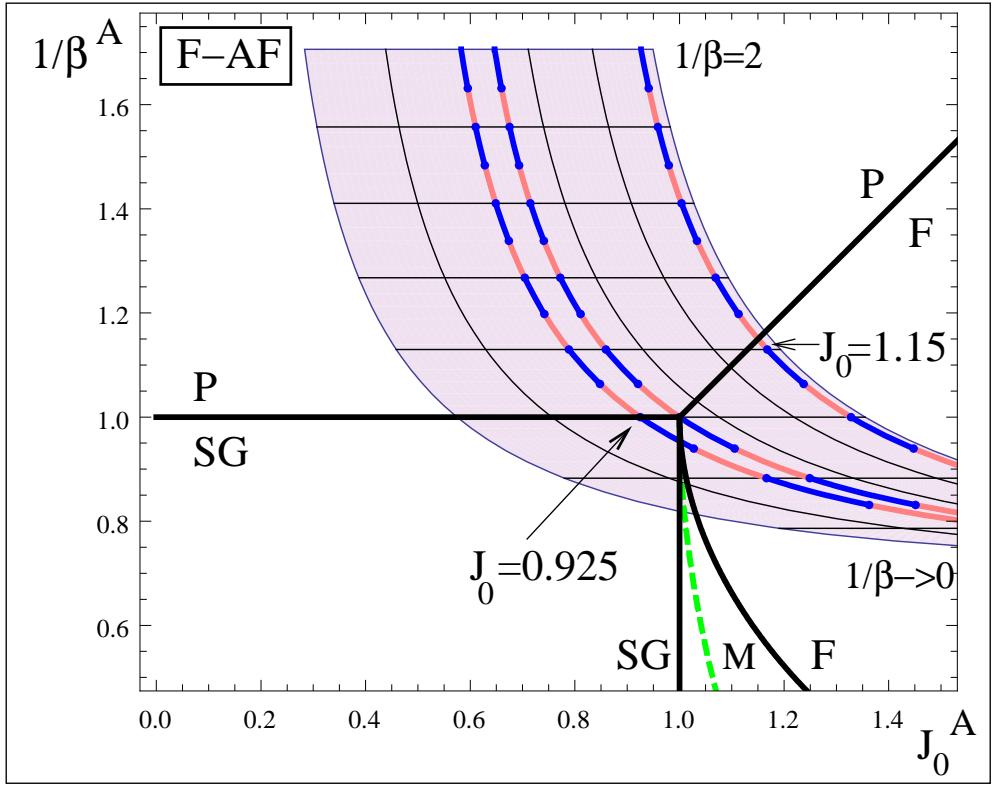}
\caption[Composite F-AF auxiliary system phase diagram.]{(color online) \label{fig:F-AF} The F-AF models (\ref{eq:composite.HamAux1}) in a parameter range ($\gamma=[0.75,1.25],1/\beta=(0,2]$) are mapped through (\ref{eq:composite.AuxiliaryMapping}) to auxiliary SK models parameterized by ($J^A_0/J^A$,$1/\beta^A$). These models are equivalent about the high temperature transition lines, and elsewhere equivalent when constraining higher than second order moments to zero (\ref{eq:composite.GENOPexpanded}). Horizontal isobars indicate constant $\beta$, and the near vertical isobars indicate constant $\gamma$, in the composite model parameter space. The set of transition lines for the SK model are shown, the upper most solid lines describing the high temperature phase transition. The SK auxiliary model predicts that as temperature is lowered in the composite models behavior converges towards a mean field ferromagnetic behavior. For small $\gamma$ the prediction is that a spin glass phase transforms through a mixed phase to an RS ferromagnet behavior as temperature is lowered. Decreasing temperature about the triple point ($\gamma=1$) there is only an RS ferromagnetic behavior. The three highlighted isobars correspond to composite systems from left to right parameterized by $\gamma=0.952$ ($J_0^{A}=0.925$ at $\beta_C$), $\gamma=1$ ($J_0^{A}=1$ at $\beta_C$) and $\gamma=1.23$ ($J_0^{A}=1.15$ at $\beta_C$), across a range of temperatures. }
}
\end{figure*}

The SK auxiliary model can be used to predict trends as temperature or some other parameter is varied in the F-AF model about the high temperature transition points. Using the mapping (\ref{eq:composite.AuxiliaryMapping}) combined with an exact (FRSB) description of the transitions and phases of the SK model at high and low temperature, the trajectories implied by the mapping can be used as a leading order predictor of phase behavior.

Choosing the F-AF models (\ref{eq:composite.HamAux1}) such that
\begin{equation}
\mB = \gamma \;; \qquad J^S = \atanh(1/\sqrt{C})\;; \label{eq:composite.mB1}
\end{equation}
a class of models parameterized by $\gamma \in [0,\infty)$ is created. The disorder in couplings decreases with $\gamma$ from a typical spin glass set to an ordered ferromagnetic set. These models are characterized by a high temperature spin glass transition at $\beta_C=1$ when $\gamma<1$, and a high temperature ferromagnetic transition at a temperature $\beta_C^{-1}=\gamma$ when $\gamma>1$. There is a triple point in the parameter space at $\gamma=1,\beta=1$. Phase transitions between ferromagnetic and spin glass phases are possible where $\beta\gtrsim 1$ and $\gamma\sim 1$.

Near the triple point model parametrization ($\gamma=1$) a decrease in temperature results in a competition between ferromagnetic and spin glass solutions. A graphical answer to which solution dominates is provided by figure~\ref{fig:F-AF}, for a range of high temperature transition properties. If only leading order moments are considered in the free energy then all composite systems evolve towards an RS ferromagnetic behavior with decreasing temperature. Thus unusual transitions away from FRSB spin-glass phases
towards RS ferromagnetic phases are predicted as temperature is lowered.

The auxiliary model is an approximation except very close to the high temperature transition, where higher order moments are negligible and linear approximations apply. At lower temperature complicated couplings with these higher order moments may prevent an FRSB to RS transition. However, working at the level of linearisation near the transition point the unusual FRSB to RS transitions are still observed in some models.

\cut{In the vicinity of the triple point the prediction of a spin-glass to ferromagnetic transition is accurate at leading order about the high temperature transition. The derivative describing the line of RS instability in the SK model is strictly vertical at the triple point, whereas the trajectory of the composite model in the auxiliary model space (\ref{eq:composite.partialJ0ApartialbetaA}) is positive as temperature is lowered. Therefore some models exhibit a transition towards first an RSB spin-glass phase with decreasing temperature, then towards an RS ferromagnetic phase; this does not preclude transitions back to RSB at lower temperature.}

In the F-AF model a spin glass phase with zero magnetization can not be a sufficient description at low temperature. This is because the spins disconnected from the sparse sub-structure can evolve independently and undergo an independent phase transition induced by the dense sub-structure. The results at leading order are in agreement with this observation.

\subsection{The AF-F model}
\label{composite.AFFauxiliarysystem}
\begin{figure*}[!htbp]
\centering{
 \includegraphics[width=0.8\linewidth]{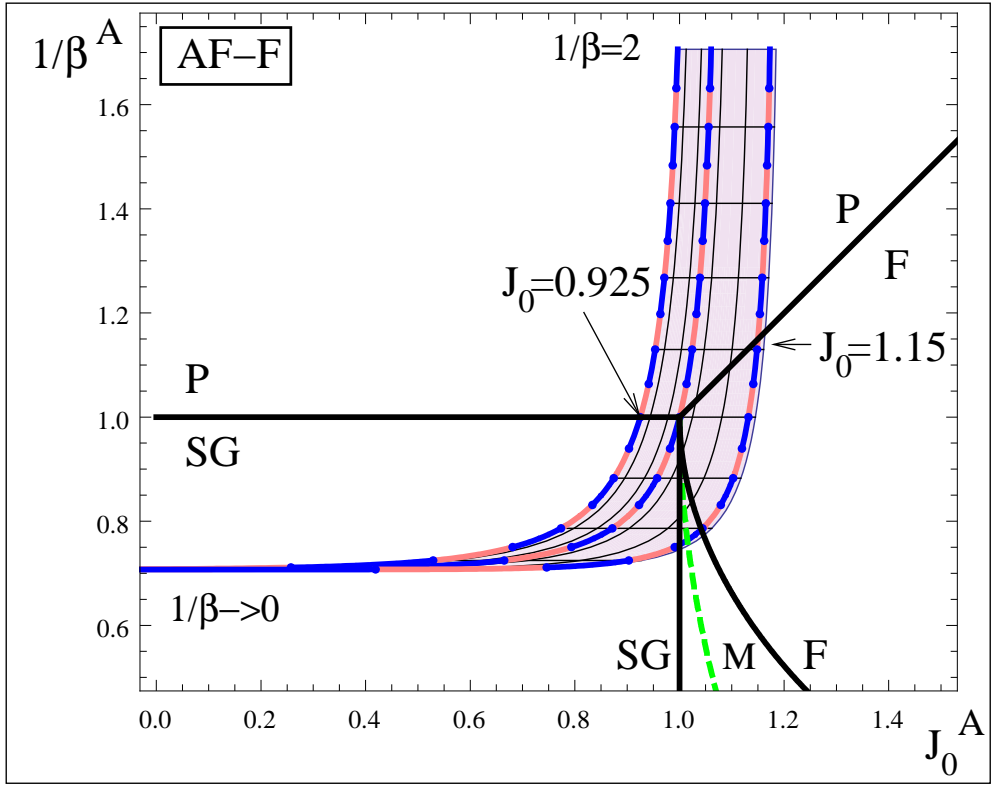}
\caption[Composite AF-F auxiliary model phase diagram.]
{(color online) \label{fig:AF-F}. The AF-F model (\ref{eq:composite.HamAux2}) as parameterized in $\gamma-\beta$ space ($\gamma=[0.75,1.25],1/\beta=(0,2]$) is mapped (\ref{eq:composite.AuxiliaryMapping}) to an auxiliary dense model parameter space. The auxiliary model prediction is that the magnetic order parameter ($m^2$) goes to zero in all composite models as temperature is lowered, a FRSB spin glass phase describes the zero temperature limit. The three highlighted systems correspond to systems with $\gamma=0.746$ ($J_0^{A}=0.925$ at $\beta_C$), $\gamma=1$ ($J_0^{A}=1$ at $\beta_C$) and $\gamma=1.23$ ($J_0^{A}=1.15$ at $\beta_C$).}
}
\end{figure*}

Consider the assignments
\begin{equation}
\mB = \gamma (1 - C \tanh(J_S/\gamma))\;, \qquad J^S = \atanh(1/\sqrt{C}) \label{eq:composite.mB2}\;,
\end{equation}
as applied to the AF-F model (\ref{eq:composite.HamAux2}), with $\gamma \in [0,J_S/\atanh\left(1/C\right)]$ describing the level of order in couplings. Larger $\gamma$ can be considered, but these correspond to systems with small ferromagnetic couplings in the dense part rather than anti-ferromagnetic ones.

The predictions based on a leading order representation of the order parameters are shown in figure~\ref{fig:AF-F}. Composite systems are predicted to evolve towards spin glass
phases as temperature is lowered; lowering temperature at large $\gamma$ results first in transitions to a stable RS ferromagnetic phases then towards a mixed phase before finally a spin glass phase. The auxiliary model predicts that at lower temperature the magnetic moment is suppressed, for all $\gamma$ up to the maximum value $J_S/\atanh\left(1/C\right)$, so that in the low temperature limit all systems are in a phase equivalent to a "finite temperature" spin-glass phase in the SK model. As
temperature is lowered RS states become unstable towards RSB, which is the scenario normally observed in dense or sparse spin glass models.

The prediction that all systems converge towards a finite temperature spin glass is a consequence of the limited moment description. The spin glass behavior is a residual effect of the sparse couplings, and at low temperature depends strongly on higher order moments which are absent in the auxiliary model. The spin glass phase is not induced by the dense anti-ferromagnetic couplings.
\subsection{Regular connectivity models}
\label{composite.DHTT}
\cut{\begin{figure}[!htbp]
\begin{center}
\includegraphics[width=\linewidth]{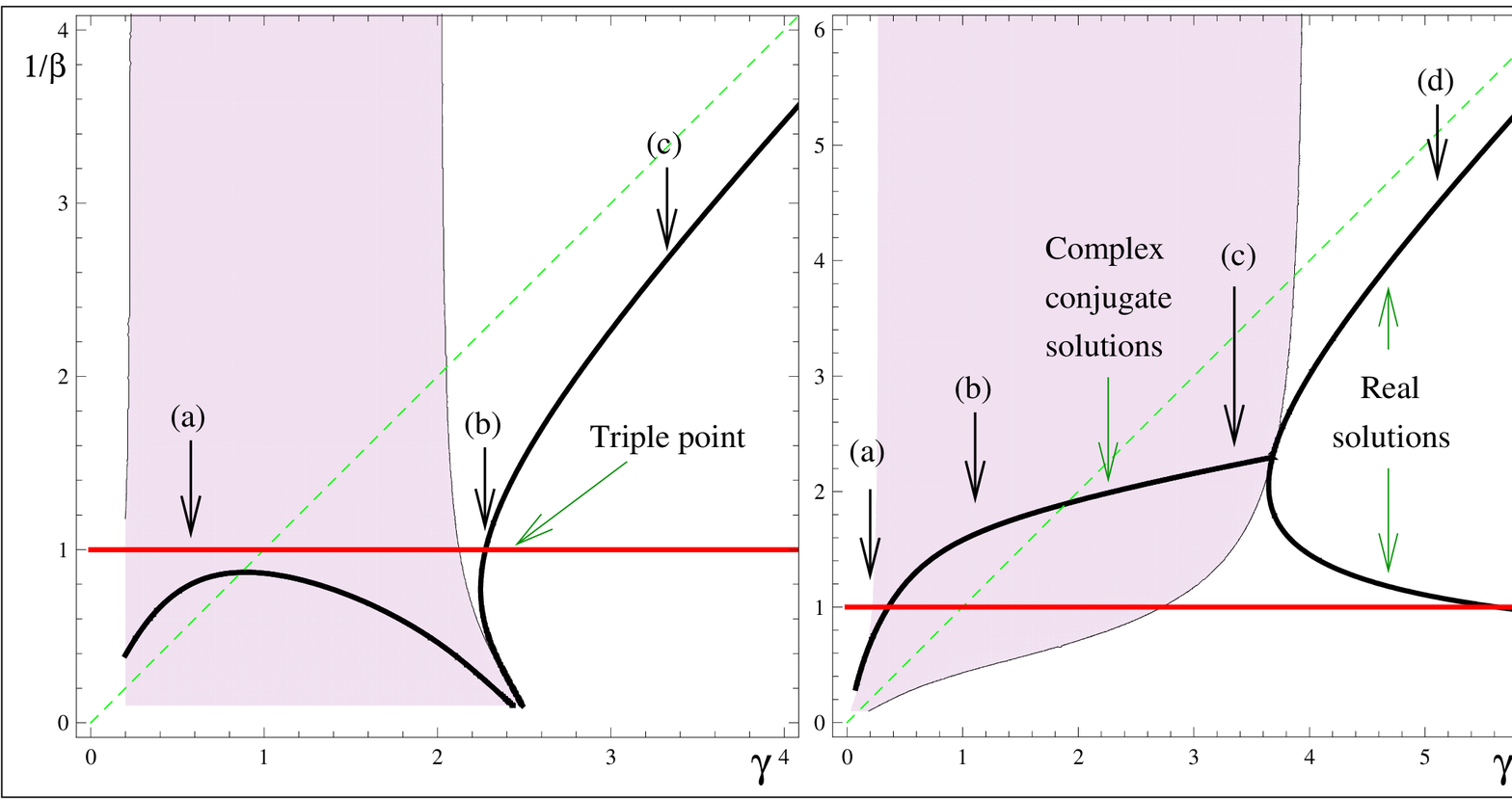}
\caption[Regular F-AF and AF-F composite model phase diagrams.] {(color online) \label{fig:composite.Regularcomptransitions} The assumption that real eigenvectors describe the dominant perturbations towards 1-spin order about the paramagnetic solution breaks down for some regular connectivity composite models (shaded region). The horizontal thick line indicate the P-SG instability, below which a spin glass solution emerges continuously. The dashed diagonal line indicates the P-F instability under an
assumption that the mean and variance of couplings strengths describe the transition (\ref{eq:composite.reglimHTT}). The solid curved line indicates the points in parameter space where the modulus of the eigenvectors (\ref{eq:composite.HTT3}) is one, below which
a stable ferromagnetic solution may exist. Left figure: In the AF-F model continuous high temperature transitions are expected everywhere. (a) A spin glass transition is found. (b) A triple point is observed, with a single dominated ferromagnetic orientation described by a real principal eigenvector $(\qbal,\qal)$. (c) Continuous ferromagnetic transitions are observed at large $\gamma$. Under the mapping (\ref{eq:composite.mBreg}), 
only systems up to $\gamma\sim 2.5$, just beyond the triple point, are valid AF-F models. Larger $\gamma$ describe ferromagnetic rather than anti-ferromagnetic dense couplings. Right figure: In the F-AF model a discontinuous transition to the ferromagnetic phase may be anticipated for intermediate $\gamma$. (a) The high temperature transition is to a spin glass phase. (b) The modulus (\ref{eq:composite.HTT3}) exceeds one, and so a solution in competition with the paramagnetic phase is expected, but a ferromagnetic
solution does not appear continuously. The paramagnetic solution is unstable first towards a spin-glass phase as temperature is lowered. (c) A discontinuous transition from a paramagnetic phase to ferromagnetic phase is anticipated, with metastability in a range of parameters. (d) With strong dense ferromagnetic couplings a continuous high temperature transition is observed towards an RS ferromagnetic state.}
\end{center}
\end{figure}
}
\begin{figure}[!htbp]
\begin{center}
\includegraphics[width=\linewidth]{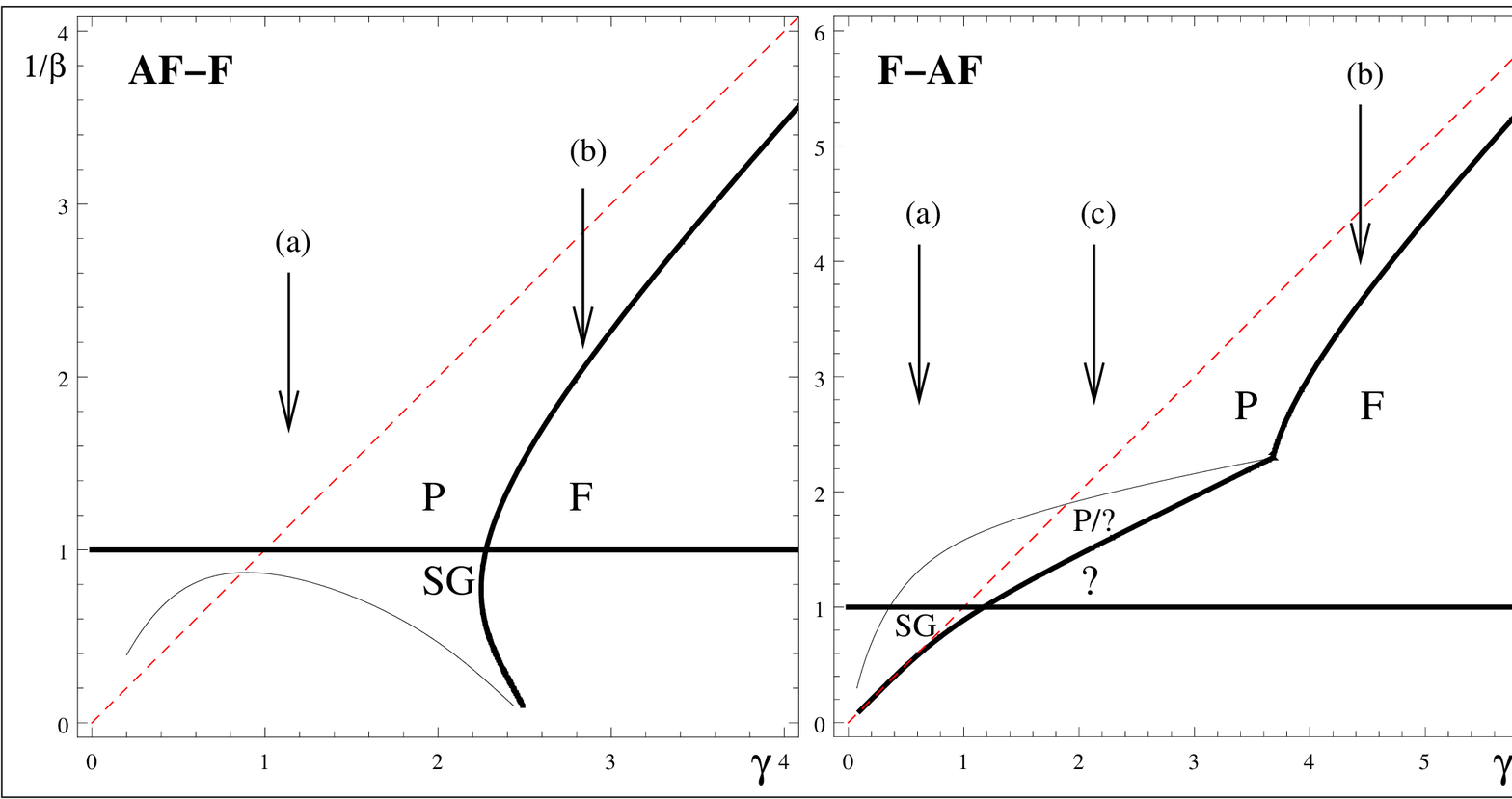}
\caption[Regular F-AF and AF-F composite model phase diagrams.]
{(color online) \label{fig:composite.Regularcomptransitions}
This figure demonstrates the critical points in the linear analysis of the paramagnetic solution for regular connectivity models.
In this figure the horizontal line indicates a high temperature instability in the paramagnetic solution towards two spin order. Other lines indicate instabilities towards 1-spin order: the straight diagonal line is assuming the conditions~(\ref{eq:composite.reglimHTT}), the thick and thin lines are the points where the real part or modulus of the principal eigenvector(s) equal one, respectively. Left figure: In the AF-F model decreasing temperature results in either a continuous  spin glass or ferromagnetic transition. (a) At small $\gamma$ a spin glass phase emerges continuously with decreasing temperature. (b) At large $\gamma$ eigenvectors describing 1-spin order are real, a continuous ferromagnetic transition is found.  Right figure: In the F-AF  model continuous and discontinuous transitions occur, no continuous transition triple-point exists. (a) At small $\gamma$ eigenvectors describing the 1-spin order are complex, but a spin glass high temperature transition is dominant. (b) At large $\gamma$ a continuous transition occurs described by a real eigenvector. (c) An instability in the paramagnetic solution in the first moment is anticipated at the lower (thick) line for intermediate $\gamma$, the properties of the discontinuously  emerging solution (labeled ?) cannot be established by a linearized approach. The thin line indicates instability in the modulus for the linearized system, which is speculated to relate to the existence of 
the non-paramagnetic solution.
}
\end{center}
\end{figure}

Figure~\ref{fig:composite.Regularcomptransitions} demonstrates the limitations on the parameter range consistent with unique locally stable RS solutions, in the case of regular connectivity systems. The two figures correspond to the models (\ref{eq:composite.HamAux1}) and (\ref{eq:composite.HamAux2}), but with regular couplings (\ref{eq:composite.HamAux4}) and mean connectivity $3$. The coupling scaling is
\begin{equation}
\mB = \gamma (1 - C \tanh(J_S/\gamma)) \;; \qquad J^S = \atanh(1/\sqrt{C-1}) \label{eq:composite.mBreg}\;.
\end{equation}
The choice of $J^S$ ensures that everywhere temperature $\beta=1$ corresponds to a spin glass instability in the paramagnetic solution. The choice of scaling means that under the approximated ferromagnetic transition scheme (\ref{eq:composite.reglimHTT}), the critical temperature implying local instability in the paramagnetic solution towards ferromagnetism increases linearly with $\gamma$, denoted by the dashed line in
figure~\ref{fig:composite.Regularcomptransitions}. If the transition were predicted by (\ref{eq:composite.reglimHTT}) then a triple point would occur at $1$: for $\gamma<1$ all high temperature transitions would be of a spin glass type; and for $\gamma>1$ transitions would be of a ferromagnetic type.

With couplings characterized by (\ref{eq:composite.mBreg}),
a range of $\gamma$ values allow the instability of the paramagnetic
solution towards one-spin order to be described by complex eigenvalues.
In the AF-F regular model,
at small values of $\gamma$, the complex eigenvectors describe the stability of the paramagnetic solution towards 1-spin order, but as temperature is lowered a spin glass instability is first attained.
At larger $\gamma$ (equivalently $J_0$) a triple point is reached,
but here the eigenvectors are real, and a continuous transition from the
paramagnetic to ferromagnetic phase is found at larger $\gamma$.

In the F-AF regular model complex eigenvalues occur in a parameter range relevant to the high temperature transition. When $J_0$ is sufficiently large a continuous high temperature ferromagnetic transition is observed, and at small $\gamma$ there is a continuous spin-glass transition. There exists a broad range of $\gamma$ between these regimes where the ferromagnetic solution can not emerge continuously from the paramagnetic solution and two locally stable solutions are anticipated. There is no triple-point in this model suitable for a perturbative analysis.

In a small number of Metropolis-Hasting Monte-Carlo simulations~\cite{Landau:GMC} two attractors corresponding to paramagnetic and ferromagnetic type configurations were found in these parameter ranges, though no systematic analysis was undertaken.

\section{Replica symmetric solution of low temperature behavior}
\label{composite.lowtemp}

In figures~\ref{fig:J085},\ref{fig:J1} and~\ref{fig:J115} stability measures and magnetizations for the composite models, equivalent at $\beta_C$ to SK models with
 $J_0^{A}\!=\!1$, $J_0^{A}\!=\!1.15$ and $J_0^{A}=0.925$, are presented at various temperatures below the $1/\beta_C$. The trends found are compared to those predicted by the auxiliary model in the vicinity of the transitions, as shown in figures~\ref{fig:F-AF} and~\ref{fig:AF-F}, and also RS solutions to dense (SK) and sparse (VB) models with equivalent high temperature properties.

\subsection{Numerical evaluation of the saddle-point equations}
To work beyond a perturbative approach the RS saddle-point equations are solved by population dynamics~\cite{Mezard:BLSG}. The results are presented based on samples from a single run of a population dynamics algorithm. In population dynamics machine numbers are used for $m$ and $q$ and the distribution $\RSOP$ is represented by an order-parameter histogram ($\Histogram$) of $N$ components
\begin{equation}
\RSOP \rightarrow \Histogram = \left\lbrace h_1, \ldots, h_N \right\rbrace \label{eq:composite.Histogram}\;.
\end{equation}
The saddle-point equations (\ref{eq:composite.RSsaddle1})-(\ref{eq:composite.RSsaddle3}) are treated as a mapping with integrals and summations replaced by random samples. This implies a random map from the histogram to itself. Updating Histograms recursively by a large number of random maps, from a random initial condition, leads to an accurate description of the fixed point $\RSOP$.

The random sampling is done in such a way as to reduce fluctuations in the variance of the Gaussian distributed samples, and mean of the Poissonian distributed samples, to $O(1/N)$. A single iteration includes an update of every field in the histogram $\Histogram$ with either parallel or random sequential order. Given that anti-ferromagnetic couplings play a role in the dynamics, there is a risk that an invalid macroscopic anti-ferromagnetic state could be amplified by parallel updates. This
scenario does not form a problematic point in the analysis undertaken, but was relevant to work undertaken in~\cite{Raymond:OC}, and carefully avoided. In order to control finite size effects a scheme of microcanonical sampling was employed with respect to $\Histogram$, so that each field in generation $(t)$ is involved in forming exactly $C$ fields in generation $(t+1)$.

A histogram of $65556$ floating point fields run for $1024$ iterations appears to resolve all statistical quantities of interest down to a temperature of $\sim 1/(10 \beta_C)$, with great precision, even in the vicinity of phase transitions. At lower temperature there is a rapid decrease in the resolution of statistical quantities, which is uniform across tested systems and probably related to numerical precision limitations in the representation for hyperbolic functions. Based on the converged set of order parameters samples are taken in the following $256$ iterations to determine robust system statistics.

The initial condition for the order parameters $m^2$, $q$ and $\Histogram$ are chosen as paramagnetic, combined with a small systematic bias towards spin-glass and ferromagnetic configurations with small, but non-zero values to the dense sub-structure moments ($m^2=q$), elements of $\Histogram$ are sampled according to a Gaussian ${\cal N}(m,q)$ such that the mean and variance of the histogram values are $m+O(1/N)$ and $q=O(1/N)$. Other initial conditions were also tested to ensure that dynamical bias was not implied by initial conditions, the suggested scheme converged effectively and systematically.

\subsubsection{Numerical evaluation of the stability equations}

The longitudinal stability is tested by initialising a fluctuation histogram $\delta \Histogram$
\begin{equation}
\delta \Histogram = \{ (\chi^2)^{(t)}_1, (\chi^2)^{(t)}_2,
\ldots,(\chi^2)^{(t)}_N \}\;,
\end{equation}
where each component corresponds to a distinct field in the histogram $\Histogram$ (\ref{eq:composite.Histogram}). Each component represents a topology free measure of $\bar{\delta h^2}_{i \rightarrow j}^{(t)}$, each of which is evolved according to (\ref{eq:composite.VarianceProp}), with the site dependent fields and parameters replaced by a sample of fields from $\Histogram$ and other quenched disorder determined as in the
field update. Cases in which $J^2=0$ ($q^{(t)}=0$), without linear perturbations are considered. The stability exponent is
\begin{equation}
\lambda^{(t)} = \log \frac{ \sum_l (\chi^2)_l^{(t)}}{ \sum_l (\chi^2)_l^{(t-1)}} \label{eq:composite.lambda}\;,
\end{equation}
which is negative if BP is convergent in expectation. This is averaged over many generations, alongside renormalization of $\delta \Histogram$ to prevent numerical precision problems.

\subsection{The F-AF and AF-F models}
\begin{figure*}
\centering{
 \includegraphics[width=0.8\linewidth]{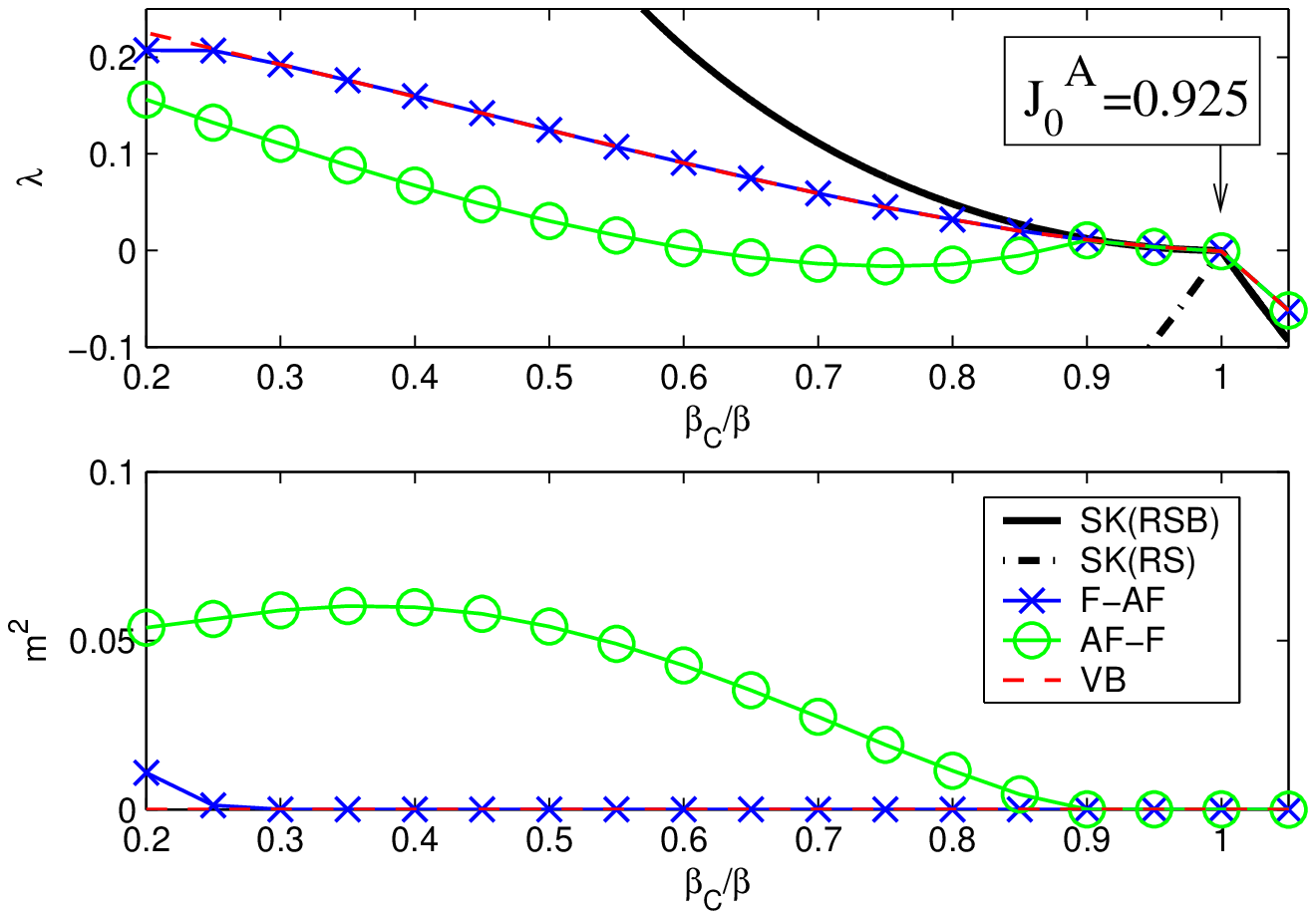}
\caption[Composite system behavior lowering temperature from a SG-P phase transition.]
{(color online) \label{fig:J085} A comparison of the stability exponent and magnetization for the F-AF (circles),AF-F (crosses), VB (dashed line) and SK (solid line) models under the RS assumption. Every model is equivalent at the high temperature spin glass transition point 
to an SK model parameterized by $J^{A}_0=0.925$, and temperature variation is considered on the rescaled interval $\beta_C/\beta=[0.2,1.05]$. In the top figure two stability exponents are given for the SK model, a longitudinal measure SK(RS) and a latitudinal measure SK(RSB). In the lower figure the sparse and dense models show similar trends with $\lambda>0$ and $m^2=0$. Composite models behave as sparse spin-glass models whenever $m^2=0$, but there is a departure in both models at low temperature. In all models as
temperature decreases $\lambda>0$, except for the F-AF model which is negative over an intermediate temperature range. Both composite models attain a non-zero magnetic moment at low temperature, which is not seen in the VB or SK models. The F-AF model is in approximate agreement with figure~\ref{fig:F-AF} at high temperature. However, the behaviors observed in the composite models at low temperature are not anticipated by the auxiliary model.}
}
\end{figure*}

\begin{figure*}
\centering
 \includegraphics[width=0.8\linewidth]{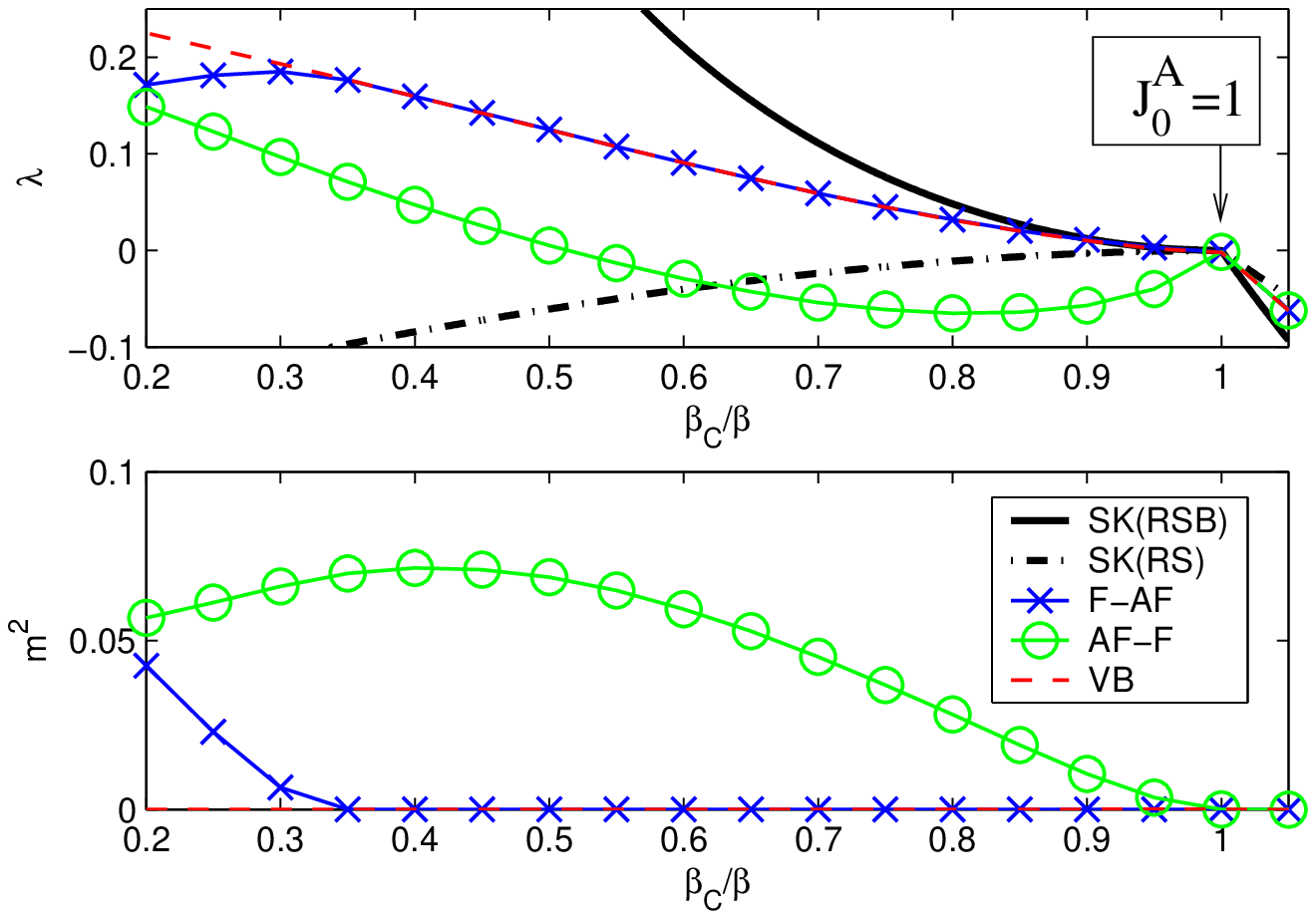}
\caption[Composite system behavior lowering temperature from a triple-point.]
{(color online) \label{fig:J1} A comparison of the longitudinal stability and magnetization for the F-AF (circles),AF-F (crosses), VB (dashed line) and SK (solid line) models under the RS assumption. Every model is equivalent at the high temperature transition to an SK model with $J^{A}_0=1$, coincident with the triple point in the phase diagram. Temperature variation is considered on the rescaled interval $\beta_C/\beta=[0.2,1.05]$. Trends differ in F-AF from figure~\ref{fig:F-AF} in that the magnetization acquires a maximum value, and the stability exponent tends towards a positive value at sufficiently low temperatures. Trends differ in AF-F from figure~\ref{fig:AF-F} in the appearance of a magnetic moment at low temperatures.}
\end{figure*}

\begin{figure*}
\centering
 \includegraphics[width=0.8\linewidth]{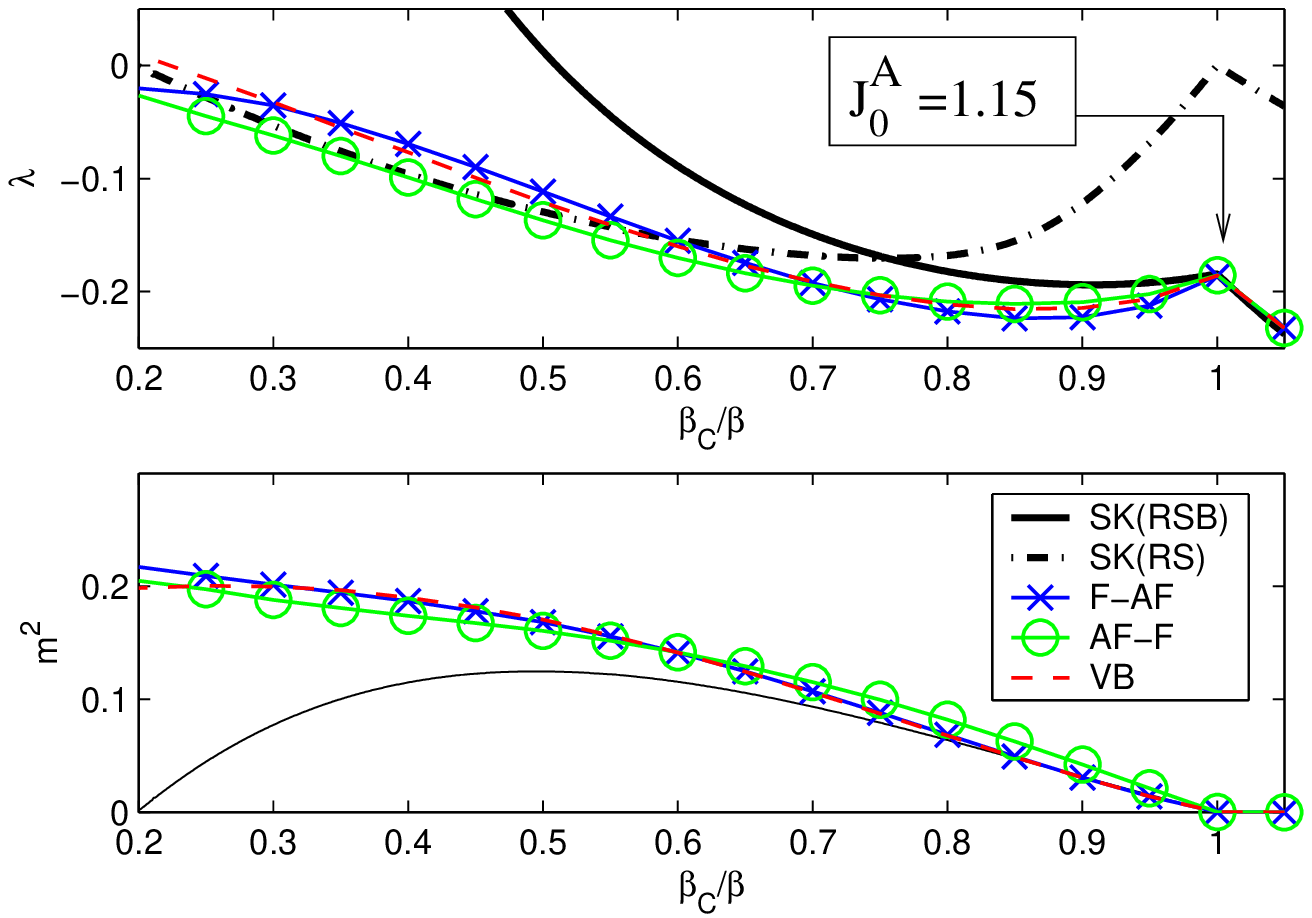}
\caption[Composite system behavior lowering temperature from a F-P phase transition.]{(color online) \label{fig:J115} A comparison of the longitudinal stability and magnetization for the F-AF (circles),AF-F (crosses), VB (dashed line) and SK (solid line) models under the RS assumption. Every model is equivalent at the high temperature ferromagnetic transition point to an SK model with $J^{A}_0=1.15$, and temperature variation is considered on the rescaled interval $\beta_C/\beta=[0.2,1.05]$. Two stability exponents are given for SK. The marginal stability at the
paramagnetic-ferromagnetic transition point ($\beta_C/\beta=1$) is with respect to a linear instability, which is captured by the longitudinal instability exponent [SK(RS)], but not by the other non-linear stability exponents. Properties of the F-AF model display features of the VB model rather than the auxiliary model predictions (figure~\ref{fig:F-AF}). Trends also differ in AF-F model from figure~\ref{fig:AF-F}, as instability is not realized until much lower than the predicted temperature, properties are again closer to the VB model.}
\end{figure*}

Results for VB, SK, F-AF and AF-F models are shown. The VB model presented for comparison is of connectivity $2$, the same as the sparse sub-structures for F-AF and AF-F models, and has a balance of anti-ferromagnetic and ferromagnetic interactions described by a $\pm J$ model (\ref{eq:composite.pmJ}).
Figures~\ref{fig:J085}, \ref{fig:J1} and \ref{fig:J115} follow the dashed lines in figures~\ref{fig:F-AF} and \ref{fig:AF-F} to highlight the behavior of the system as temperature is lowered along fixed $J_0^{A}$ values. Figure~\ref{fig:J085} demonstrates the results for the set of systems equivalent at the high temperature transition point to a dense model with $J_0^{A}=0.925$.  In all systems there is a high temperature transition that is $P-SG$ at $\beta_C=1$, behavior is examined for relative temperature $\beta_C/\beta$ in the interval $(0.1,1.05)$.

The stability exponent ($\lambda$) and magnetization ($m^2$) are identical in all the models very close to the transition, the phase is a spin glass ($m=0$,$q>0$) and the RS description is unstable ($\lambda>0$). The F-AF model becomes unstable towards a mixed (unstable RS ferromagnetic) phase at relatively high temperature. This is qualitatively similar to the prediction based on the auxiliary model of the composite system (see figure (\ref{fig:F-AF}), and the transition temperature is comparable to what would be predicted by the auxiliary model.

When the magnetization is zero (the spin glass solution) only the even moments of the
distribution in the composite models contribute to their behavior. These include only sparse model dependent parts for F-AF, AF-F so that these models are described by a saddle-point solution identical to the sparse model.

In the AF-F model the ferromagnetic order parameter is suppressed down to a temperature $\beta_C/\beta \approx 0.25$ where it acquires a small value. This is close to the point where $q$ reaches a maximum value, saturation is reached before $q=1$ due to the disconnected component in the sparse sub-structure. This low temperature transition must have a strong dependence on higher order moments since it is in strong contrast with the auxiliary model prediction (figure~\ref{fig:AF-F}).

Figure~\ref{fig:J1} demonstrates results for the same models and temperature range, but for cases in which the models have a high temperature triple-point transition. In this
figure the F-AF model has a behavior that is clearly distinct from the other three models. As temperature is lowered a ferromagnet phase is found rather than a spin glass phase in the other cases, in agreement with figure~\ref{fig:F-AF}. At lower temperatures a maximum magnetization is reached and a small decrease in magnetization is discernable at the lowest values in the temperature range. With $\beta_C/\beta<0.5$ the RS ferromagnetic phase becomes unstable to a mixed phase.

Initially, at high temperatures, the AF-F model is described by a spin glass phase. With the continuous emergence of a ferromagnetic moment at low temperature there is a decrease in the stability exponent.

In figure~\ref{fig:J115} the behavior of systems exhibiting a high temperature ferromagnetic transition are shown, systems with auxiliary models defined by $J_0^{A}=1.15$ at the high temperature transition. In this regime reentrant behavior is seen in the SK model, but not in the VB or composite models. The two composite models follow very closely the behavior of the VB model, although at $\beta_C/\beta\sim 0.3$ there appears to be a modification of the trend in the stability exponent for the AF-F model absent in the F-AF and VB models.

The ferromagnetic moment is largest in the AF-F model at high temperature, and the F-AF model at low temperatures. There are also several such cross overs in the stability exponent. The RS solutions are stable for the composite systems and VB over the full temperature range presented.

\section{Reentrant behavior and structure in finite systems}
\label{composite.finite}
\subsection{BP and Monte-Carlo simulation}
\begin{figure}[htb]
\begin{center}
\includegraphics[width=\linewidth]{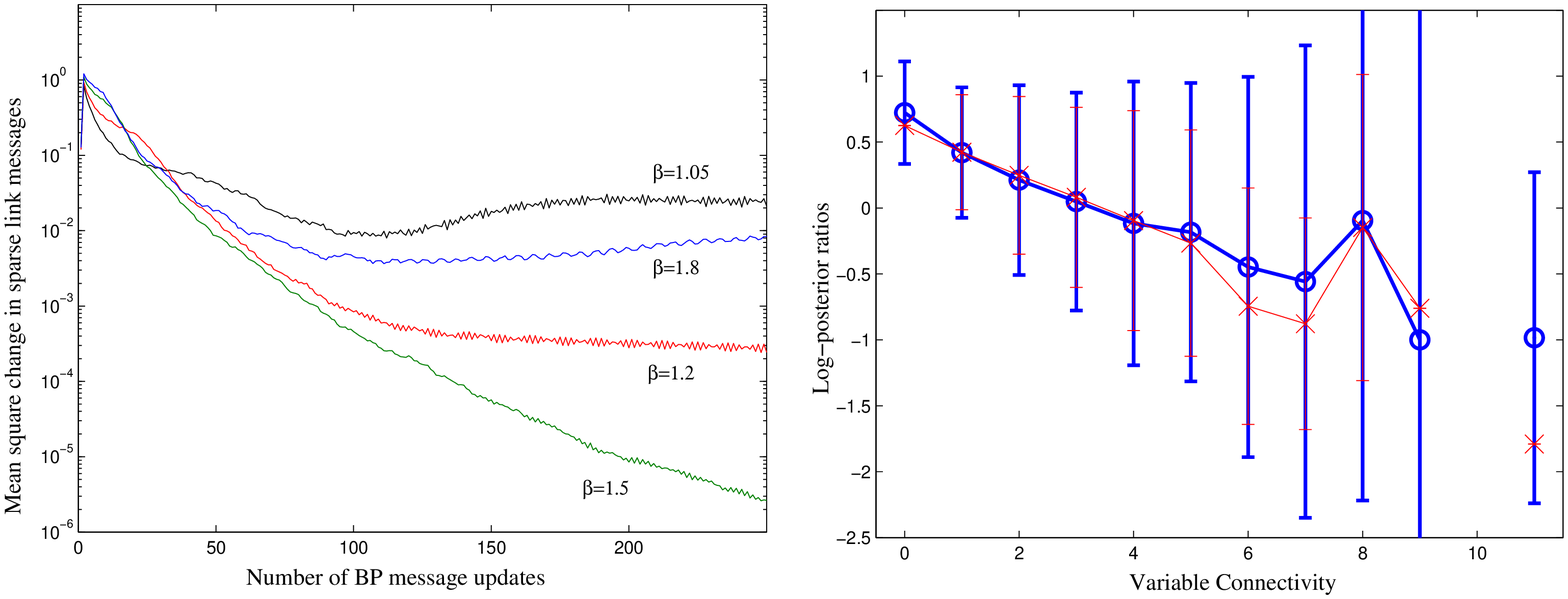}
\caption[Reentrant stability behavior in BP and monte-carlo simulations.]
{(color online) \label{fig:composite.BPresult} Results in applying BP and Monte-Carlo simulation to an F-AF model of size $5000$ spins, and $\gamma=1$ for various temperatures. Left figure: Iteration of BP on a sample graph from various initial conditions is convergent for this sample of quenched disorder at intermediate temperature only, as indicated by the
exponential decay in the stability measure. Right: For the case where BP converges $\beta=1.5$, the mean and variance in the field distribution are demonstrated as a function of variable connectivity in the sparse sub-structure. Thick lines (circles) demonstrate the results of Metropolis-Hastings Monte-Carlo simulation. Thin lines (crosses) demonstrate the estimates of BP. These are in agreement except at high variable connectivity. The magnetism of the system is supported by the alignment of low connectivity variables, with variables of high connectivity in the sparse sub-structure
being magnetized in the opposite direction.}
\end{center}
\end{figure}

Some testing of thermodynamic results was undertaken in samples of $N \!= \!O(100)-\!\!O(8000)$ spins by sampling through a Metropolis-Hastings
algorithm~\cite{Landau:GMC}, and estimating log-posterior ratios by BP. These studies verified qualitatively the outcomes of the thermodynamic analysis at high temperature. The paramagnetic phase was observed to transform continuously into either a ferromagnetic, spin glass, or mixed (unstable ferromagnetic) phase as temperature was decreased. The ferromagnetic state is assumed to be described by a connected phase space up to finite size effects. Stability of the BP algorithm was measured through the mean square change in BP log-posterior estimates (\ref{eq:composite.H})
\begin{equation}
\lambda^{(t)} = \frac{1}{N} \sum_{i=1}^N \left(H_i^{(t)}-H_i^{(t+1)}\right)^2 \;,
\end{equation}
this being a new definition of $\lambda$ related to (\ref{eq:composite.lambda}), but distinguished by the algorithmic context.

Figure~\ref{fig:composite.BPresult} demonstrates a simulation of an F-AF model with $5000$ spins. This demonstrates that the non-monotonic behavior seen in the RS solution of the F-AF model, and predicted by the leading order expansion, can be realized in finite systems also. The second part of the figure demonstrates the structure of the magnetic phase in the F-AF model. The macroscopic magnetization is supported primarily by spins coincident with the disconnected component in the sparse sub-structure.

\subsection{Structure of phases and transitions}

In the F-AF model the inhomogeneity in magnetizations, with the disconnected component being the most strongly aligned set of variables, seems an intuitive and necessary feature in a model with such a stark contrast in coupling types.

The disconnected component appears to play an even more vital role in the AF-F model. In the magnetic phase of this model all the disconnected components are observed in Monte-Carlo and BP experiments to be anti-correlated with the macroscopic magnetization, which is an intuitive result. Whereas almost all other variables, connected through the sparse sub-structure take values aligned with the macroscopic order. In the large system limit there should be some discrimination in the topology within the sparse-substructure. Some important topological features of sparse Poissonian graphs are outlined in figure~\ref{fig:composite.2cores}.  In general the highly connected spins may take one alignment, the disconnected component an opposite alignment, with other variables intermediate.

The inhomogeneity in the structure must also be vital in allowing continuous transitions between various phases, and in the dynamics of models. The continuous emergence of a magnetic phase as temperature is lowered in the AF-F model is presumably by a nucleation process, whereas in the F-AF model the ferromagnetic part can emerge first in the disconnected component and percolate inwards to the core of the sparse sub-structure. The absence of sufficient inhomogeneity in the regular connectivity models is responsible for the metastability found in some parameter ranges.


\section{Summary}

We have investigated composite models that combine well-studied disordered spin systems with densely and sparsely interacting components with the expectation that the combined model would exhibit new feature that have not been observed in the original systems. Here we focused on the case of competing interactions of ferromagnetic and anti-ferromagnetic nature and have explored the phase diagrams for different interaction types and relative strengths. The equilibrium properties were examined through the replica method, exhibiting a new reentrant behavior from spin glass to ferromagnetic phases as temperature is lowered, and transitions from replica symmetry broken to replica symmetric phases. A discontinuous transition from the paramagnetic to ferromagnetic phases has been observed in regular connectivity graphs.

While the models investigated here comprise classical and well-understood models, they provide insight and new understanding of simple complex systems that combine different structures, and different levels and interaction types. Such systems have been increasingly the subject of interdisciplinary research activities using methods of varying mathematical rigor. We believe that established methods of statistical physics are highly suitable for understanding both macroscopic and microscopic properties of such systems and that the current investigation paves the way for the study of similar complex systems.

\begin{acknowledgments}
Jack Raymond is appreciative of the advice given in improving this document by Juan Neirotti and Ton Coolen. Support from EPSRC grant EP/E049516/1 is gratefully acknowledged.
\end{acknowledgments}

\UPDATEBIBYES{
\bibliographystyle{apsrev}
\bibliography{BibliographyMAY09_v4}

\begin{thebibliography}{25}
\expandafter\ifx\csname natexlab\endcsname\relax\def\natexlab#1{#1}\fi
\expandafter\ifx\csname bibnamefont\endcsname\relax
  \def\bibnamefont#1{#1}\fi
\expandafter\ifx\csname bibfnamefont\endcsname\relax
  \def\bibfnamefont#1{#1}\fi
\expandafter\ifx\csname citenamefont\endcsname\relax
  \def\citenamefont#1{#1}\fi
\expandafter\ifx\csname url\endcsname\relax
  \def\url#1{\texttt{#1}}\fi
\expandafter\ifx\csname urlprefix\endcsname\relax\def\urlprefix{URL }\fi
\providecommand{\bibinfo}[2]{#2}
\providecommand{\eprint}[2][]{\url{#2}}

\bibitem[{\citenamefont{Fischer and Hertz}(1991)}]{Fischer:SG}
\bibinfo{author}{\bibfnamefont{K.}~\bibnamefont{Fischer}} \bibnamefont{and}
  \bibinfo{author}{\bibfnamefont{J.}~\bibnamefont{Hertz}},
  \emph{\bibinfo{title}{Spin Glasses}} (\bibinfo{publisher}{Cambridge
  University Press}, \bibinfo{address}{Cambridge, UK}, \bibinfo{year}{1991}).

\bibitem[{\citenamefont{M\'{e}zard et~al.}(1987)\citenamefont{M\'{e}zard,
  Parisi, and Virasoro}}]{Mezard:SGT}
\bibinfo{author}{\bibfnamefont{M.}~\bibnamefont{M\'{e}zard}},
  \bibinfo{author}{\bibfnamefont{G.}~\bibnamefont{Parisi}}, \bibnamefont{and}
  \bibinfo{author}{\bibfnamefont{M.}~\bibnamefont{Virasoro}},
  \emph{\bibinfo{title}{Spin Glass Theory and Beyond}}
  (\bibinfo{publisher}{World Scientific}, \bibinfo{address}{Singapore},
  \bibinfo{year}{1987}).

\bibitem[{\citenamefont{Hertz et~al.}(1991)\citenamefont{Hertz, Krogh, and
  Palmer}}]{Hertz:ITNC}
\bibinfo{author}{\bibfnamefont{J.}~\bibnamefont{Hertz}},
  \bibinfo{author}{\bibfnamefont{A.}~\bibnamefont{Krogh}}, \bibnamefont{and}
  \bibinfo{author}{\bibfnamefont{R.}~\bibnamefont{Palmer}},
  \emph{\bibinfo{title}{Introduction to the theory of neural computation}}
  (\bibinfo{publisher}{Addison-Wesley}, \bibinfo{address}{Boston, MA, USA},
  \bibinfo{year}{1991}).

\bibitem[{\citenamefont{Richardson and Urbanke}(2008)}]{Richardson:MCT}
\bibinfo{author}{\bibfnamefont{T.}~\bibnamefont{Richardson}} \bibnamefont{and}
  \bibinfo{author}{\bibfnamefont{R.}~\bibnamefont{Urbanke}},
  \emph{\bibinfo{title}{Modern Coding Theory}} (\bibinfo{publisher}{Cambridge
  University Press}, \bibinfo{address}{Cambridge, UK}, \bibinfo{year}{2008}).

\bibitem[{\citenamefont{Nishimori}(2001)}]{Nishimori:SP}
\bibinfo{author}{\bibfnamefont{H.}~\bibnamefont{Nishimori}},
  \emph{\bibinfo{title}{Statistical Physics of Spin Glasses and Information
  Processing}} (\bibinfo{publisher}{Oxford Science Publications},
  \bibinfo{address}{Oxford, UK}, \bibinfo{year}{2001}).

\bibitem[{\citenamefont{Bollobas}(2001)}]{Bollobas:RG}
\bibinfo{author}{\bibfnamefont{B.}~\bibnamefont{Bollobas}},
  \emph{\bibinfo{title}{Random Graphs}} (\bibinfo{publisher}{Cambridge
  University Press,Cambridge}, \bibinfo{address}{Cambridge, UK},
  \bibinfo{year}{2001}), \bibinfo{edition}{2nd} ed.

\bibitem[{\citenamefont{Sherrington and Kirkpatrick}(1975)}]{Sherrington:SMSG}
\bibinfo{author}{\bibfnamefont{D.}~\bibnamefont{Sherrington}} \bibnamefont{and}
  \bibinfo{author}{\bibfnamefont{S.}~\bibnamefont{Kirkpatrick}},
  \bibinfo{journal}{Phys. Rev. Lett.} \textbf{\bibinfo{volume}{35}},
  \bibinfo{pages}{1792} (\bibinfo{year}{1975}).

\bibitem[{\citenamefont{Ellis}(1985)}]{Ellis:ELD}
\bibinfo{author}{\bibfnamefont{R.}~\bibnamefont{Ellis}},
  \emph{\bibinfo{title}{Entropy, large deviations and statistical mechanics}}
  (\bibinfo{publisher}{Springer-Verlag}, \bibinfo{address}{New York, NY, USA},
  \bibinfo{year}{1985}).

\bibitem[{\citenamefont{Viana and Bray}(1985)}]{Viana:PD}
\bibinfo{author}{\bibfnamefont{L.}~\bibnamefont{Viana}} \bibnamefont{and}
  \bibinfo{author}{\bibfnamefont{A.}~\bibnamefont{Bray}}, \bibinfo{journal}{J.
  Phys. C} \textbf{\bibinfo{volume}{18}}, \bibinfo{pages}{3037}
  (\bibinfo{year}{1985}).

\bibitem[{\citenamefont{Mottishaw}(1987)}]{Mottishaw:RSB}
\bibinfo{author}{\bibfnamefont{P.}~\bibnamefont{Mottishaw}},
  \bibinfo{journal}{Europhys. Lett.} \textbf{\bibinfo{volume}{4}},
  \bibinfo{pages}{333} (\bibinfo{year}{1987}).

\bibitem[{\citenamefont{Skantzos and Coolen}(2000)}]{Skantzos:1ID}
\bibinfo{author}{\bibfnamefont{N.~S.} \bibnamefont{Skantzos}} \bibnamefont{and}
  \bibinfo{author}{\bibfnamefont{A.~C.~C.} \bibnamefont{Coolen}},
  \bibinfo{journal}{J. Phys. A} \textbf{\bibinfo{volume}{33}},
  \bibinfo{pages}{5785} (\bibinfo{year}{2000}).

\bibitem[{\citenamefont{Raymond and Saad}(2009)}]{Raymond:CC}
\bibinfo{author}{\bibfnamefont{J.}~\bibnamefont{Raymond}} \bibnamefont{and}
  \bibinfo{author}{\bibfnamefont{D.}~\bibnamefont{Saad}}, \bibinfo{journal}{J.
  Stat. Mech.} \textbf{\bibinfo{volume}{2009}}, \bibinfo{pages}{P05015 (25pp)}
  (\bibinfo{year}{2009}).

\bibitem[{\citenamefont{Hase and Mendes}(2008)}]{Hase:DA}
\bibinfo{author}{\bibfnamefont{M.~O.} \bibnamefont{Hase}} \bibnamefont{and}
  \bibinfo{author}{\bibfnamefont{J.~F.~F.} \bibnamefont{Mendes}},
  \bibinfo{journal}{J. Phys. A} \textbf{\bibinfo{volume}{41}},
  \bibinfo{pages}{145002 (9pp)} (\bibinfo{year}{2008}).

\bibitem[{\citenamefont{Albert et~al.}(2000)\citenamefont{Albert, Jeong, and
  Barabasi}}]{Albert:EA}
\bibinfo{author}{\bibfnamefont{R.}~\bibnamefont{Albert}},
  \bibinfo{author}{\bibfnamefont{H.}~\bibnamefont{Jeong}}, \bibnamefont{and}
  \bibinfo{author}{\bibfnamefont{A.}~\bibnamefont{Barabasi}},
  \bibinfo{journal}{Nature} \textbf{\bibinfo{volume}{406}},
  \bibinfo{pages}{378} (\bibinfo{year}{2000}).

\bibitem[{\citenamefont{Raymond and Saad}(2008)}]{Raymond:OC}
\bibinfo{author}{\bibfnamefont{J.}~\bibnamefont{Raymond}} \bibnamefont{and}
  \bibinfo{author}{\bibfnamefont{D.}~\bibnamefont{Saad}}, \bibinfo{journal}{J.
  Phys. A} \textbf{\bibinfo{volume}{41}}, \bibinfo{pages}{324014 (30pp)}
  (\bibinfo{year}{2008}).

\bibitem[{\citenamefont{Monasson}(1998)}]{Monasson:OP}
\bibinfo{author}{\bibfnamefont{R.}~\bibnamefont{Monasson}},
  \bibinfo{journal}{J. Phys. A} \textbf{\bibinfo{volume}{31}},
  \bibinfo{pages}{513} (\bibinfo{year}{1998}).

\bibitem[{\citenamefont{M\'{e}zard and Parisi}(2001)}]{Mezard:BLSG}
\bibinfo{author}{\bibfnamefont{M.}~\bibnamefont{M\'{e}zard}} \bibnamefont{and}
  \bibinfo{author}{\bibfnamefont{G.}~\bibnamefont{Parisi}},
  \bibinfo{journal}{Eur. Phys. Jour. B} \textbf{\bibinfo{volume}{20}},
  \bibinfo{pages}{217} (\bibinfo{year}{2001}).

\bibitem[{\citenamefont{Kschischang et~al.}(2001)\citenamefont{Kschischang,
  Frey, and Loeliger}}]{Kschischang:FG}
\bibinfo{author}{\bibfnamefont{F.}~\bibnamefont{Kschischang}},
  \bibinfo{author}{\bibfnamefont{B.}~\bibnamefont{Frey}}, \bibnamefont{and}
  \bibinfo{author}{\bibfnamefont{H.-A.} \bibnamefont{Loeliger}},
  \bibinfo{journal}{IEEE Trans. on Info. Theory} \textbf{\bibinfo{volume}{47}},
  \bibinfo{pages}{498} (\bibinfo{year}{2001}).

\bibitem[{\citenamefont{Kabashima}(2003)}]{Kabashima:PB}
\bibinfo{author}{\bibfnamefont{Y.}~\bibnamefont{Kabashima}},
  \bibinfo{journal}{J. Phys. Soc. Jpn.} \textbf{\bibinfo{volume}{72}},
  \bibinfo{pages}{1645} (\bibinfo{year}{2003}).

\bibitem[{\citenamefont{Mallard and Saad}(2008)}]{Mallard:BPDG}
\bibinfo{author}{\bibfnamefont{E.}~\bibnamefont{Mallard}} \bibnamefont{and}
  \bibinfo{author}{\bibfnamefont{D.}~\bibnamefont{Saad}},
  \bibinfo{journal}{Phys. Rev. E} \textbf{\bibinfo{volume}{78}},
  \bibinfo{pages}{021107} (\bibinfo{year}{2008}).

\bibitem[{\citenamefont{Rivoire et~al.}(2004)\citenamefont{Rivoire, Biroli,
  Martin, and M\'{e}zard}}]{Rivoire:GM}
\bibinfo{author}{\bibfnamefont{O.}~\bibnamefont{Rivoire}},
  \bibinfo{author}{\bibfnamefont{G.}~\bibnamefont{Biroli}},
  \bibinfo{author}{\bibfnamefont{O.}~\bibnamefont{Martin}}, \bibnamefont{and}
  \bibinfo{author}{\bibfnamefont{M.}~\bibnamefont{M\'{e}zard}},
  \bibinfo{journal}{Eur. Phys. Jour. B} \textbf{\bibinfo{volume}{37}},
  \bibinfo{pages}{55} (\bibinfo{year}{2004}).

\bibitem[{\citenamefont{Almeida and Thouless}(1978)}]{Almeida:SSK}
\bibinfo{author}{\bibfnamefont{J.~d.} \bibnamefont{Almeida}} \bibnamefont{and}
  \bibinfo{author}{\bibfnamefont{D.}~\bibnamefont{Thouless}},
  \bibinfo{journal}{J. Phys. A} \textbf{\bibinfo{volume}{11}},
  \bibinfo{pages}{983} (\bibinfo{year}{1978}).

\bibitem[{\citenamefont{Mottishaw and De~Dominicis}(1987)}]{Mottishaw:SRF}
\bibinfo{author}{\bibfnamefont{P.}~\bibnamefont{Mottishaw}} \bibnamefont{and}
  \bibinfo{author}{\bibfnamefont{C.}~\bibnamefont{De~Dominicis}},
  \bibinfo{journal}{J. Phys. A} \textbf{\bibinfo{volume}{20}},
  \bibinfo{pages}{L375} (\bibinfo{year}{1987}).

\bibitem[{\citenamefont{Landau and Binder}(2005)}]{Landau:GMC}
\bibinfo{author}{\bibfnamefont{D.}~\bibnamefont{Landau}} \bibnamefont{and}
  \bibinfo{author}{\bibfnamefont{K.}~\bibnamefont{Binder}},
  \emph{\bibinfo{title}{A Guide to Monte Carlo Simulations in Statistical
  Physics}} (\bibinfo{publisher}{Cambridge University Press},
  \bibinfo{address}{Cambridge, UK}, \bibinfo{year}{2005}),
  \bibinfo{edition}{2nd} ed.

\bibitem[{\citenamefont{Raymond}(2008)}]{Raymond:Thesis}
\bibinfo{author}{\bibfnamefont{J.}~\bibnamefont{Raymond}}, Ph.D. thesis,
  \bibinfo{school}{Aston University}, \bibinfo{address}{Birmingham, UK}
  (\bibinfo{year}{2008}).

\end{thebibliography}
}
\UPDATEBIBNO{

}
\appendix

\section{Replica Calculation}
\label{app:CompositeSystem_Replica}

Carrying out the calculation via the replica method involves a combination of sparse and dense quenched disorder averages. It is convenient to define the sparse substructure in terms of an adjacency matrix $\mA$: labeling each edge by $\mu$  and  each variable $k$, $A_{\mu k}=\{0,1\}$. With mean connectivity $C$ the number of edges is $C N/2$ so that in the absence of other constraints, the probability distribution is defined
\begin{equation}
P(\mA) = \prod_{\mu=1}^{C N/2} \left[{N \choose 2}^{-1} \delta\left(\sum_k A_{\mu k} - 2 \right) \right]\;.
\end{equation}
This is a micro-canonical description of interactions, but formulations with the number of edges not strictly fixed (to $C N/2$) are possible. In the limit of large $N$ this describes a Poissonian distribution in the variable connectivity. Both Poissonian and regular connectivity are sufficiently described in typical case analysis by
\begin{equation}
P(\mA) \propto \prod_{\mu=1}^{C N/2} \left[\frac{1}{2} \delta\left(\sum_k A_{\mu k} - 2 \right)\right] \prod_{i=1}^{N} \< \frac{c_f\factorial}{C^{c_f}} \delta\left(\sum_\mu A_{\mu k} - {c_f} \right) \>_{c_f}\prod_{\mu,k} P(A_{\mu k}) \;,\label{appeq:composite.factorgraphconnectivitymatrix}
\end{equation}
the average in $c_f$ being with respect to the marginal variable connectivity distribution of mean $C$~\cite{Raymond:Thesis} and taking $P(A_{\mu k})$ to be a sparse prior
\begin{equation}
P(A_{\mu k}) = \left(1- \frac{2}{N}\right)\delta_{A_{\mu k}} + \frac{2}{N}\delta_{A_{\mu k},1}\;.
\end{equation}

The Hamiltonian may be written in a form
\begin{equation}
\Ham = \frac{1}{2} \sum_\mu J^S_\mu \left[ \left( \sum_k A_{\mu k} \tau_k \right)^2 - 2\right] + \sum_{\ij} J^D_\ij S_i S_j \;,\label{appeq:composite.Ham}
\end{equation}
where the representation of the dense part is unmodified from (\ref{eq:composite.Ham}),
$J^S_\mu$ is the random sparse coupling sampled according to $\phi(x)$ in (\ref{eq:composite.phix}), but can be replaced by the integration variable $x$ in the self averaged expressions. The replicated partition function is
\begin{equation}
\begin{array}{lcl}
\repZ &=& \prod_\alpha\left[\sum_{\vS^\alpha}\right] \<\prod_\mu \<\exp\left\lbrace \frac{\beta}{2} x \sum_\alpha \left[\left(\sum_k A_{\mu k} S^\alpha_k \right)^2-2\right] \right\rbrace \>_{x} \>_{\mA} \\
&\times& \prod_\ij \<\exp \left\lbrace \beta J^D_\ij \sum_\alpha S_i^\alpha S_j^\alpha \right\rbrace\>_{J^D_\ij}\;.
\end{array}
\end{equation}
Since the Hamiltonian is factorized with respect to the sparse and dense quenched variables, these averages may be taken independently.

In the sparse part it is useful to linearize the squared components with a Hubbard-Stratonovich transform for each factor node and replica index pair
\begin{equation}
\< \cdots \>_{\mA} = \int \prod_{\mu,\alpha} \left[ \rmD\lambda_\mu^\alpha\right] \< \prod_\mu \<\exp \left\lbrace-\beta x n \right\rbrace \prod_k \left[\exp \left\lbrace \sqrt{\beta x} \sum_\alpha \lambda_\mu^\alpha S^\alpha_k \right\rbrace \right]^{A_{\mu k}}\>_{x} \>_{\mA}\;, \label{appeq:composite.sparsepart}
\end{equation}
with
\begin{equation}
\int \rmD \lambda = \frac{1}{\sqrt{2\pi}} \int \rmd \lambda \exp\left\lbrace -\frac{\lambda^2}{2} \right\rbrace\;.
\end{equation}

The delta functions in the adjacency matrix probability distribution (\ref{appeq:composite.factorgraphconnectivitymatrix}) can be represented by complex contour integrals
\begin{equation}
\delta\left(L_\mu \!-\! \sum_k A_{\mu k} \right) = \prod_\mu \intY{2} \prod_{k} Y_\mu^{A_{\mu k}}\;;\qquad  \delta\left(C_k \!-\! \sum_\mu A_{\mu k} \right) = \prod_k \intZ{C_k}\prod_{\mu} Z_k^{A_{\mu k}}\;,
\end{equation}
using the notation
\begin{equation}
\oint \rmD_x X = \frac{1}{2\pi \rmi} \oint \frac{\rmd X}{X^{x}} \;;\qquad \oint \rmD_{\vx} \vX = \prod_z \oint \rmD_{x_z} X_z\;.
\end{equation}
The contours can be taken around the origin in the complex plane, so the identities are assured by Cauchy's residue formula.

Factorizing various terms (\ref{appeq:composite.sparsepart}) becomes
\begin{equation}
\< \cdots \>_{\mA} \propto \<\oint {\rmD_\vC \vZ} \prod_\mu \< \exp \left\lbrace-\beta x n \right\rbrace \int\rmD\rvlambda \oint \rmD_{2}Y_\mu  \prod_k \left[\sum_{A_{\mu k}} P(A_{\mu k}) \arbitraryfunction\right]\>_{x} \>_{\vC}\;, \label{appeq:composite.sparsepart2}
\end{equation}
where $\rmD\rvlambda \equiv \prod_{\mu,\alpha} \rmD_{\mu}^{\alpha}$, $\arbitraryfunction$ abbreviates a number of terms factorized in $k$. Subject to three integrals and an average on connectivities the dependence on the adjacency matrix is factorized in $\mu$  and $k$, and the trace over $\mA$ can be completed
\begin{equation}
\prod_k \left[\sum_{A_{\mu k}} P(A_{\mu k}) \arbitraryfunction\right] = \prod_k \left[\left(1-\frac{2}{ N} \right) + \frac{2}{N} Y_\mu Z_k\exp \left\lbrace \sqrt{\beta x} \sum_\alpha \lambda_\mu^\alpha S^\alpha_k \right\rbrace\right] \label{appeq:composite.sparsepart3}
\end{equation}
Taking the integral in $Y_\mu$ picks out only the residue, the second term in the expansion, so that
\begin{equation}
\prod_k \left[\sum_{A_{\mu k}} P(A_{\mu k}) \arbitraryfunction\right] \propto \sum_\ij \prod_{l=\{i,j\}} \left[\frac{1}{N} Z_l \exp \left\lbrace \sqrt{\beta x} \sum_\alpha \lambda_\mu^\alpha S^\alpha_l \right\rbrace \right]\;.\label{appeq:composite.sparsepart4}
\end{equation}
Introducing an identity
\begin{equation}
1 = \sum_\rvsigma \prod_\alpha \delta_{\sigma^\alpha,\sigma^\alpha_k}\;,
\end{equation}
and extracting the $k$ dependence through the order parameter definition
\begin{equation}
1 = \int \prod_{\rvsigma} \left[ \rmd \GENOP(\rvsigma) \delta\left(\GENOP(\rvsigma) - \frac{1}{N} \sum_k Z_k \delta_{\rvsigma,\rvsigma_k}\right) \right]\;,
\end{equation}
all quenched variable dependence is factorized except in the order parameter definition. In principle the identity applies to the entire complex plane, but relevant part of the order parameter is assumed to be real, the imaginary part takes a value $0$ in the final saddle-point formulation by assumption. The results are self-consistent given this assumption, and the necessity of real valued order parameters can be demonstrated in some special cases (see Appendix~\ref{app:ConjugateFields}).

Having taken the average in $\mA$ the Hubbard-Stratonovich transform may be inverted to give up to ensemble dependent constants
\begin{equation}
\begin{array}{lcl}
\< \cdots \>_{\mA} &=& \int \prod_{\rvsigma}\rmd \GENOP(\rvsigma) \prod_{k}\< \frac{C_k!}{C^{C_k}} \intZ{{C_k}} \prod_{\rvsigma} \delta\left(\GENOP(\rvsigma) - \frac{1}{K} \sum_k Z_k \delta_{\rvS_k,\rvsigma} \right)\>_{C_k} \\
&\times& \prod_\mu \left[ \sum_{\rvtau,\rvsigma} \GENOP(\rvtau)\GENOP(\rvsigma)\<\exp \left\lbrace \beta x \sum_\alpha \tau^\alpha \sigma^\alpha \right\rbrace \>_{x} \right] \end{array}\;,
\end{equation}
where the average is with respect to the sparse coupling distribution (\ref{eq:composite.phix}).

The dense part of the Hamiltonian can be expanded to second order allowing averages in the quenched couplings
\begin{equation}
\<\cdots\>_{\mJ^D} = \prod_\ij \left[ 1 + \beta \frac{J_0}{N} \sum_\alpha S_i^{\alpha} S_j^{\alpha} + \frac{\beta^2 J^2}{2 N} \sum_\alal S_i^{\alpha_1} S_i^{\alpha_2} S_j^{\alpha_1} S_j^{\alpha_2} \right]\;.
\end{equation}
Defining the dense order parameters
\begin{equation}
\qal = \frac{1}{N} \sum_i S^\alpha_i \;;\qquad \qalal = \frac{1}{N} \sum_i S^{\alpha_1}_i S^{\alpha_2}_i \;,
\end{equation}
the $k$ dependence is extracted
\begin{equation}
\begin{array}{lcl}
\<\cdots\>_{\mJ^D} &=& \prod_\alpha \delta\left(q_\alpha - \frac{1}{N}\sum_k S_k^\alpha\right) \prod_\alal \delta\left(\qalal - \frac{1}{N}\sum_k S_k^{\alpha_1} S_k^{\alpha_2}\right) \\
&\times& \exp \left\lbrace \frac{N \beta J_0}{2} \sum_\alpha q_\alpha^2\right\rbrace \exp \left\lbrace \frac{N\beta^2 J^2}{2} \sum_\alal q_{\alal}^2\right\rbrace
\end{array}
\end{equation}
The definitions of $\qal$, $\qalal$ and $\GENOP$, introduced as $\delta$-functions may be Fourier transformed introducing conjugate parameters, for the real part of $\GENOP(\rvsigma)$
\begin{equation}
\delta\left(\GENOP(\rvsigma) - \frac{1}{N} Z_k\sum_k \delta_{\rvsigma,\rvsigma_k}\right) = \int_{\rmi\infty}^{-\rmi\infty} \exp\left\lbrace - C N \GENOPconj(\rvsigma)\GENOPconj \right\rbrace \exp\left\lbrace - C \sum_k Z_k \GENOPconj(\rvsigma_k)\right\rbrace
\end{equation}
By either an explicit calculation of contour integrals, or by a self-consistent assumption, the integral on the complex line (the standard definition of the Fourier transform) might be considered to be rotated onto the real line so that real valued conjugate parameters can be considered.

The scaling of the Fourier transform by $N$ reflects an assumption of extensive entropy and is also a necessary feature in scalable solutions of the saddle-point method. With the specific choice the order parameter $\GENOP$ is normalized. The choice of an additional factor $C$ in the sparse order parameter definition is chosen so that $\GENOPconj$ and $\GENOP$ are normalized.

The trace over replicated spins is finally taken to give an expression for free energy (\ref{eq:composite.replicafreeenergy}), composed of terms (\ref{eq:composite.G1}), (\ref{eq:composite.G3}) and ($\Gtwo$) up to ensemble dependent constant terms and $O(1/N)$ corrections. The $\Gtwo$ term is (\ref{eq:composite.G2}) in the Poissonian variable connectivity, and in general
\begin{equation}
\Gtwo \!=\! - \log \<\sum_{\rvS} \left( \GENOPconj(\rvS) \right)^{c_f} \exp\left\lbrace \!\sum_\alpha \qhal S_\alpha \!+\! \sum_{\alal} \qhalal S^{\alpha_1} S^{\alpha_2} \right\rbrace \>_{{c_f}} \label{appeq:composite.G2gen}\;,
\end{equation}
where the average over ${c_f}$ is with respect to the marginal variable connectivity distribution, uniform or Poissonian.

\subsection{Modifications to the saddle-point equations}
The saddle-point equations can be written down for the general case (\ref{appeq:composite.G2gen}), the generalization of (\ref{eq:composite.saddle1}) in the sparse order parameter is
\begin{equation}
\GENOP(\rvsigma) \propto \<c_f \left[\GENOPconj(\rvsigma)\right]^{c_f-1}\exp \left\lbrace \sum_\alpha \qhal \sigma^\alpha + \sum_\alal \qhalal \sigma^{\alpha_1}\sigma^{\alpha_2}\right\rbrace\>_{c_f} \label{appeq:composite.GENOP_saddle_gen}\;.
\end{equation}
The dense order parameters are determined through the recursions
\begin{equation}
\qal = \sum_\rvsigma \sigma_\alpha \localreplicaprobability(\rvsigma) \;; \qquad \qalal = \sum_\rvsigma \sigma^{\alpha_1} \sigma^{\alpha_2} \localreplicaprobability(\rvsigma) \label{appeq:composite.q_saddle_gen} \;;
\end{equation}
subject to a normalized distribution
\begin{equation}
\localreplicaprobability(\rvS) = \<\left[\GENOPconj(\rvS)\right]^{c_f} \exp\left\lbrace \!\sum_\alpha \qhal S_\alpha \!+\!
\sum_{\alal} \qhalal S^{\alpha_1} S^{\alpha_2}
\right\rbrace \>_{{c_f}}\;.
\end{equation}
The conjugate saddle-point equations are unchanged in form (\ref{eq:composite.saddle2}).

\section{Order parameter considerations}
\label{app:ConjugateFields}
An interpretation for some parameters can be gained by consideration of derivatives of the free energy with respect to $\beta$ and simple random external fields $\vrandomfield$. This may also be used to prove the consistency of some method assumptions in the case of replica symmetry. The choice of a random field is primarily to allow a concise inclusion within the replica method equations. It is equivalent to working directly with fields that are conjugate to quantities such as $\sum_\ij \tau_i \tau_j$, or with annealed random fields in some cases.

A derivative of $\beta \safed$, with respect to $\beta$ determines the energy density. It is well known the entropy becomes negative in both the VB and SK models at low temperature if incorrect symmetry assumptions are used, this is also observed in some of the systems presented. Consider also a perturbation on the Hamiltonian to include a non-zero external field
\begin{equation}
\Delta\Ham(\vtau) = \sum_i z_i \tau_i \;;\qquad z_i = z \eta_i \label{appeq:composite.standardfields}\;;
\end{equation}
where $\eta_i$ is either uniform or randomly sampled from a symmetric distribution, and $z$ is small and positive. In the case that $\eta_i$ is uniform, the derivative of the free energy density with respect to $z$ in the limit of small $z$ is easy to evaluate and gives the magnetization, which is coincident with the sum over first replica moments at the saddle-point in the self-averaging case
\begin{equation}
m = \left.\frac{\partial}{\partial n}\right|\sum_\alpha q^*_\alpha = \frac{1}{N}\sum_i \<\tau_i\>\;.\label{appeq:magnetization}
\end{equation}
The derivative when $\veta=\{\vones,-\vones\}$ is coincident with the susceptibility and also the sum over 2-spin correlations at the saddle-point
\begin{equation}
 \Susceptibility = \left.\frac{\partial}{\partial n}\right|_{n=0}\sum_\alal q^*_\alal = \frac{1}{N}\sum_\ij \<\tau_i \tau_j\> - \<\tau_i\>\<\tau_j\> \;.\label{appeq:composite.susceptibility}
\end{equation}
These are useful quantities in evaluating the emergence of ferromagnetic order, and in determining phase transitions.

\subsection{\label{app:assumptionR} Assumption of real valued integration variables}

An assumption of the saddle-point method used to evaluate the exponential term describing the free energy is that only real valued integration parameters (order-parameters) need be considered. This Appendix demonstrates that any physical solution must be real valued in its first two moments, this is assumed to extend to higher order moments in the sparse order parameter $\GENOP$.

A useful variation on (\ref{appeq:composite.standardfields}) for purposes of general analysis are identified by a class of fields aligned with interactions. Consider the factor graph representation of the Hamiltonian (\ref{appeq:composite.Ham}). Each of $O(N)$ sparse interactions also has a unique label $\mu=\ij$, with other ordered edges $\ij$ subject to weak (dense) interactions. Therefore
\begin{equation}
z_i = \eta_i \sum_{j\setminus i} \left(z^D J^D_{(i,j)} + z^S J^S_{(i,j)} \right) + \sum_{j\setminus i} \zeta_{(i,j)} \left(z^D J^D_{(i,k)} + z^S J^S_{(i,j)}\right) \label{appeq:randomfield2}\;,
\end{equation}
is well defined and includes a component dependent on the sparse substructure and one on the dense substructure. Each of $\eta_i,\zeta_{(i,j)}$ are assumed to be $0$ (a default for discussions), uniform ($1$), or quenched variables sampled independently from $\{-1,1\}$, with $\{z^S,z^D\}$ being infinitesimal real positive fields. Unordered matrices are used in (\ref{appeq:randomfield2}) to describe their ordered counterparts, so that $(i,j)$ is $\ij$ or $\<ji\>$ as ordering dictates, for each ordered pair only one quenched parameter exists.

Physical interpretation for these fields is as follows. whenever $J_0\neq 0$ and $\eta_k=1$ derivatives with respect to $z^D$ give a magnetization. When $\eta_k=\{-1,1\}$ susceptibilities such as (\ref{appeq:composite.susceptibility}) are determined. When $\zeta_{(i,j)}=1$ the derivative probes alignments of variables with couplings, again giving a measure that can distinguish an ordered phase from a paramagnetic one. The more complicated physical quantities involve a quenched random field $\zeta_{\ij}=\{-1,1\}$, for example
\begin{equation}
\left.\frac{\partial}{\partial \left[\beta(z^D)^2\right]} \right|_{z^D=0} \safed = \lim_{N\rightarrow \infty} \frac{1}{2 N} \<\left(\sum_{(i,j)} J^D_{(i,j)} S_i \right)^2\> - \frac{1}{2 N} \<\left(\sum_{(i,j)} J^D_{(i,j)} S_i \right)\>^2 \label{appeq:thisistheend}\;,
\end{equation}
identifying a type of susceptibility.

The free energy in the replica formulation, with inclusion of these infinitesimal fields involves a modification of the factor-centric ($\Gone$) term (\ref{eq:composite.G1}) in the free energy. Following Appendix~\ref{app:CompositeSystem_Replica}
\begin{equation}
\begin{array}{lcl}
\Gone &=& - \sum_\alpha \frac{1}{2}\beta J_0 (\qal + 2 z^D\<\zeta_\ij\>)\qal - \frac{1}{2} \! \sum_\alal \beta^2 J^2 \left(\qalal + 2(z^D)^2\<(\zeta_\ij)^2\>\right)\qalal \\
&-& \frac{C}{2} \log \sum_{\rvS,\rvS'} \GENOP(\rvS)\GENOP(\rvS') \int \rmd x \phi(x) \< \exp\left\lbrace - \beta x \sum_\alpha \left(S^\alpha + z^S \zeta_\ij\right) S'^\alpha \right\rbrace \>_{\zeta_\ij} \label{appeq:composite.G1}\;.\KEEPNOTE{CHECK USE OF MINUS SIGN}
\end{array}
\end{equation}
The variable-centric term is also modified from (\ref{appeq:composite.G2gen}), following from Appendix~\ref{app:CompositeSystem_Replica} the averages over $J^D_\ij$ are straightforward, but the average over $J^S_\ij$ is involved for the general case, requiring additional order parameter definitions. Consider the case that $J^S_{(i,j)}=J^S$ is uniform for simplicity. The expression then becomes
\begin{equation}
\begin{array}{lcl}
\Gtwo &=& - \log \<\sum_{\rvS} \left(\GENOPconj(\rvS) \right)^{c_f} \exp\left\lbrace \!\sum_\alpha \qhal S_\alpha \!+\! \sum_{\alal} \qhalal S^{\alpha_1} S^{\alpha_2} \right\rbrace \right.\\
&\times& \left.\exp \left\lbrace \beta \eta_k (c_f z^S + (J_0+\lambda J)z^D)\!\sum_\alpha S_\alpha \right\rbrace \>_{{c_f},\eta_k} \label{appeq:composite.G2genField}\;.
\end{array}
\end{equation}
with $\lambda$ a normally distributed parameter.

Quantities derived through the proposed choices for $\{\vzeta,\veta\}$ are necessarily real-valued at a saddle-point and may be calculated in the replica framework given the self-averaged free energy. The derivatives with respect to $\zeta_\mu$ are particularly transparent, for example taking $\eta_{\ij}$ to be a quenched parameter from $\pm 1$, the derivative at the saddle-point gives
\begin{equation}
\left.\frac{\partial}{\partial \left[\beta (z^D)^2\right]} \right|_{z^D=0} \safed = \left.\frac{\partial}{\partial n}\right|_{n=0} J^2 \sum_\alal q^*_\alal \label{appeq:keepgoingalmostthere}\;,
\end{equation}
so that whenever $J^2$ is non-zero the RS term $\sum_\alal q^*_\alal$ must be real-valued. In the case $J^2=0$ the order parameter definition is in any case redundant and can be removed from the free energy. A derivative with respect to $z^S$ can produce a similar constraint on the second moment of the sparse distribution~\cite{Raymond:Thesis}. Taking $\eta_i$ to be uniform demonstrates that first moments of the order parameter must also be real.

\subsection{Spin glass susceptibilities}
\label{app:spinglasssusceptibility}
The quantity probed through the BP stability analysis is a form of spin glass susceptibility. Consider two sets of random variables described by a joint probability distribution determined by identical quenched disorder except in a weak field term
\begin{equation}
P(\vsigma,\vtau)=\frac{1}{Z^2}\exp\left\lbrace-\Ham(\vsigma)-\Ham(\vtau) + z \sum_k \eta_k \left(z_1 \sigma_i + z_2 \sigma_j\right) \right\rbrace\;.\label{appeq:jointdistribution}
\end{equation}
where $z_1$,$z_2$ and $\eta_k$ are quenched parameters sampled independently from $\{-1,1\}$, and $z$ is a small external field. Assuming self-averaging the limit $z \rightarrow 0$ ought to smoothly recover the equilibrium description of a single system (as evaluated using an RS assumption for example). An expansion of the self-averaged free energy finds at $O(z^2)$ and $O(z^4)$ constant terms, and terms dependent on the macroscopic magnetization and susceptibility. These terms are expected to be well defined in the small $z$ limit if the population dynamics method converges, since they are coincident with the RS order parameters moments at the saddle-point, and stability is tested by fluctuations implicit in population dynamics. However, at $O(z^4)$ there is also a dependence on spin-glass susceptibility
\begin{equation}
\SpinGlassSusceptibility = \frac{1}{N} \sum_{\ij} \left(\<\sigma_i \sigma_j\> - \<\sigma_i\> \<\sigma_j\>\right)^2 \label{appeq:spinglasssusceptibility}\;.
\end{equation}
The spin glass susceptibility, as opposed to linear susceptibility (\ref{appeq:composite.susceptibility}), probes a symmetric local instability, symmetries which might allow the linear susceptibility to be convergent are absent. In the replica formulation non-convergence of the spin glass-susceptibility in the limit of small $z$ provides a sufficient criteria for failure of the RS assumption.

Non-convergence can be tested by a local stability analysis of the order parameters in a single (uncoupled) model under iteration of the saddle-point equations. The form of spin glass susceptibility tested in the BP framework is not exactly (\ref{appeq:spinglasssusceptibility}), one must consider an external field reweighed by interaction strengths to determine an appropriately reweighed set of perturbations (\ref{appeq:randomfield2}). The site factorization is then only achieved by definition of new order parameters, and it is the linear stability of the original order parameters towards this new description at $z=0$ that is a sufficient test of divergence, and comparable to stability of BP fields under iteration. However, asymptotic divergence of the BP equations as presented is expected to be a sufficient criteria for divergence of the standard spin-glass susceptibility in the equilibrium analysis, accurate as a predictor of trends and parameter dependence.

\end{document}